\documentclass[prd,twocolumn,showpacs,floatfix,amsmath,nofootinbib,amssymb,floatfix]{revtex4}
\usepackage{graphicx,color,dcolumn,booktabs,bm}
\usepackage{longtable,lscape}
\usepackage{txfonts}
\usepackage{overpic}
\usepackage{amssymb}
\usepackage{indentfirst}
\usepackage{feynmf}   
\usepackage{slashed}  
\usepackage{cases}
\usepackage{color}
\usepackage{multirow}
\usepackage{threeparttable}
\usepackage{epstopdf}
\usepackage{enumerate}
\usepackage{graphicx,color,dcolumn,booktabs,bm}
\usepackage[colorlinks, citecolor=blue,anchorcolor=red,menucolor=red, linkcolor=red,filecolor=red,urlcolor=blue,frenchlinks=red]{hyperref}

\graphicspath{{Figures/}} %

\begin{document}

\title{Low-lying charmed and charmed-strange baryon states}
\author{Bing Chen$^{1,3}$}\email{chenbing@shu.edu.cn}
\author{Ke-Wei Wei$^{1}$}\email{weikw@hotmail.com}
\author{Xiang Liu$^{2,3}$\footnote{Corresponding author}}\email{xiangliu@lzu.edu.cn}
\author{Takayuki Matsuki$^{4,5}$}\email{matsuki@tokyo-kasei.ac.jp}
\affiliation{$^1$Department of Physics, Anyang Normal University,
Anyang 455000, China\\$^2$School of Physical Science and Technology,
Lanzhou University,
Lanzhou 730000, China\\
$^3$Research Center for Hadron and CSR Physics, Lanzhou University
$\&$ Institute of Modern Physics of CAS,
Lanzhou 730000, China\\
$^4$Tokyo Kasei University, 1-18-1 Kaga, Itabashi, Tokyo 173-8602, Japan\\
$^5$Theoretical Research Division, Nishina Center, RIKEN, Saitama 351-0198, Japan}

\date{\today}

\begin{abstract}
In this work, we systematically study the mass spectra and strong decays of $1P$ and $2S$ charmed and charmed-strange baryons in the framework of nonrelativistic constituent quark models. With the light quark cluster-heavy quark picture, the masses are simply calculated by a potential model. The strong decays are studied by the Eichten-Hill-Quigg decay formula. Masses and decay properties of the well-established $1S$ and $1P$ states can be reproduced by our method. $\Sigma_c(2800)^{0,+,++}$ can be assigned as a $\Sigma_{c2}(3/2^-)$ or $\Sigma_{c2}(5/2^-)$ state. We prefer to interpret the signal $\Sigma_c(2850)^0$ as a $2S(1/2^+)$ state although at present we cannot thoroughly exclude the possibility that this is the same state as $\Sigma_c(2800)^0$. $\Lambda_c(2765)^+$ or $\Sigma_c(2765)^+$ could be explained as the $\Lambda_c^+(2S)$ state or $\Sigma^+_{c1}(1/2^-)$ state, respectively. We propose to measure the branching ratio of $\mathcal{B}(\Sigma_c(2455)\pi)/\mathcal{B}(\Sigma_c(2520)\pi)$ in future, which may disentangle the puzzle of this state. Our results support $\Xi_c(2980)^{0,+}$ as the first radial excited state of $\Xi_c(2470)^{0,+}$ with $J^P=1/2^+$. The assignment of $\Xi_c(2930)^0$ is analogous to $\Sigma_c(2800)^{0,+,++}$, \emph{i.e.}, a $\Xi^\prime_{c2}(3/2^-)$ or $\Xi^\prime_{c2}(5/2^-)$ state. In addition, we predict some typical ratios among partial decay widths, which are valuable for experimental search for these missing charmed and charmed-strange baryons.

\end{abstract}
\pacs{12.39.Jh,~13.30.Eg,~14.20.Lq} \maketitle

\section{Introduction}\label{sec1}
At present, the Particle Data Group (PDG) lists nine charmed and ten charmed-strange baryons~\cite{Agashe:2014kda}. They are $\Lambda_c(2286)^+$, $\Lambda_c(2595)^+$, $\Lambda_c(2625)^+$, $\Lambda_c(2765)^+$ (or $\Sigma_c(2765)^+$), $\Lambda_c(2880)^+$, $\Lambda_c(2940)^+$,
$\Sigma_c(2800)^{0,+,++}$, $\Sigma_c(2455)^{0,+,++}$, $\Sigma_c(2520)^{0,+,++}$,
$\Xi_c(2470)^{0,+}$, $\Xi^{\prime}_c(2580)^{0,+}$, $\Xi^\prime_c(2645)^{0,+}$, $\Xi_c(2790)^{0,+}$, $\Xi_c(2815)^{0,+}$, $\Xi^{(\prime)}_c(2930)^0$, $\Xi_c(2980)^{0,+}$, $\Xi_c(3055)^{0,+}$, $\Xi_c(3080)^{0,+}$, and $\Xi^{(\prime)}_c(3123)^+$. For brevity, we call these charmed and and charmed-strange baryons just as charmed baryons in the following.
Among these observed states, some new measurements have been performed by experiments in the past several years. The masses and widths of $\Sigma_c(2455)^{0,+,++}$, $\Sigma_c(2520)^{0,+,++}$, $\Lambda_c(2595)^+$, and $\Lambda_c(2625)^+$ have been measured with significantly small uncertainties with the efforts of CDF~\cite{Aaltonen:2011sf} and Belle~\cite{Lee:2014htd}. Very recently, the Belle Collaboration updated the measurements of $\Xi^{\prime}_c(2580)^{0,+}$, $\Xi^\prime_c(2645)^{0,+}$, $\Xi_c(2790)^{0,+}$, $\Xi_c(2815)^{0,+}$, and $\Xi_c(2980)^{0,+}$~\cite{Yelton:2016fqw}. On the other hand, new decay modes for the higher excited charmed baryon states have been found by experiments. For instance, the decay channel of $\Lambda D^+$ was first found for $\Xi_c(3055)^+$ and $\Xi_c(3080)^+$, and the following ratios of branching fractions were first reported by Belle several months ago~\cite{Kato:2016hca}:
\begin{equation}
\frac{\mathcal{B}(\Xi_c(3055)^+\rightarrow\Lambda D^+)}{\mathcal{B}(\Xi_c(3055)^+\rightarrow\Sigma_c(2455)^{++}K^-)}=5.09\pm1.01\pm0.76, \nonumber
\end{equation}
\begin{equation}
\frac{\mathcal{B}(\Xi_c(3080)^+\rightarrow\Lambda D^+)}{\mathcal{B}(\Xi_c(3080)^+\rightarrow\Sigma_c(2455)^{++}K^-)}=1.29\pm0.30\pm0.15, \nonumber
\end{equation}
and
\begin{equation}
\frac{\mathcal{B}(\Xi_c(3080)^+\rightarrow\Sigma_c(2520)^{++}K^-)}{\mathcal{B}(\Xi_c(3080)^+\rightarrow\Sigma_c(2455)^{++}K^-)}=1.07\pm0.27\pm0.01,\nonumber
\end{equation}
where the uncertainties are statistical and systematic. Obviously, these new measurements are very useful to understand the nature of these excited charmed baryon states.

Theoretically, the charmed baryons which contain one heavy quark and two light quarks occupy a particular position in the baryon physics. Since the chiral symmetry and heavy quark symmetry (HQS) can provide some qualitative insight into the dynamics of charmed baryons, the investigation of charmed baryons should be more helpful for improving our understanding of the confinement mechanism. The spectroscopy of charmed baryons has been investigated in various models. So far, the several kinds of quark potential models~\cite{Migura:2006ep,Majethiya:2008fe,Garcilazo:2007eh,Ebert:2007nw,Ebert:2011kk}, the relativistic flux tube (RFT) model~\cite{Chen:2014nyo,Chen:2009tm}, the coupled channel model~\cite{Romanets:2012hm}, the QCD sum rule~\cite{Wang:2010hs,Wang:2015kua,Chen:2015kpa}, and the Regge phenomenology~\cite{Guo:2008he} have been applied to study the mass spectra of excited charmed baryons, and so did the Lattice QCD~\cite{Padmanath:2013bla,Padmanath:2015bra}.
The strong decay behaviors of charmed baryons have been studied by several methods, such as the heavy hadron chiral perturbation theory (HHChPT)~\cite{Cheng:2006dk,Cheng:2015naa}, the chiral quark model~\cite{Zhong:2007gp,Liu:2012sj}, the $^3P_0$ model~\cite{Chen:2007xf}, and a non-relativistic quark model~\cite{Nagahiro:2016nsx}. The decays of $1P$ $\Lambda_c$ and $\Xi_c$ baryons have also been investigated by a light front quark model~\cite{Tawfiq:1998nk,Tawfiq:1999vz}, a relativistic three-quark model~\cite{Ivanov:1999bk}, and the QCD sum rule~\cite{Zhu:2000py}.

Although many experimental and theoretical efforts have been made for the research of charmed baryons, most of the $1P$ and $2S$ charmed baryons are not yet established. Several candidates, including $\Lambda_c(2765)^+$, $\Sigma_c(2800)^{0,+,++}$, $\Xi_c(2930)^0$, and $\Xi_c(2980)^{0,+}$ are still in controversy. $\Lambda_c(2765)^+$ was first observed by the CLEO Collaboration in the decay channel of $\Lambda_c(2765)^+\rightarrow\Lambda_c^+\pi^+\pi^-$~\cite{Artuso:2000xy}, and confirmed by Belle in the mode $\Sigma_c(2455)\pi$~\cite{Abe:2006rz,Joo:2014fka}. Because both $\Lambda_c^+$ and $\Sigma_c^+$ excitations can decay through $\Lambda_c^+\pi^+\pi^-$ and $\Sigma_c(2455)\pi$, we even do not know whether the observed charmed baryon signal around 2765 MeV is the $\Lambda_c^+$ or $\Sigma_c^+$ state, or their overlapping structure~\cite{Eakins:2010zz}. In the $e^+e^-$ annihilation process, an isotriplet state, $\Sigma_c(2800)^{0,+,++}$, was observed by Belle in the channel of $\Lambda_c^+\pi$, and was tentatively identified as the $\Sigma_{c2}$ state with $J^P=3/2^-$~\cite{Mizuk:2004yu}. Interestingly, another neutral resonance was later found by BaBar in the process of $B^-\rightarrow\Sigma_c^{*0}\bar{p}\rightarrow\Lambda_c^+\pi^-\bar{p}$ with the mass $2846\pm8\pm10$ MeV and decay width $86^{+33}_{-22}$ MeV~\cite{Aubert:2008ax}. The higher mass and the weak evidence of $J=1/2$ indicate that the signal observed by BaBar might be different from the Belle's observation. In this paper, we denote the signal discovered by BaBar as $\Sigma_c(2850)^0$. $\Xi_c(2930)^0$ which was only seen by BaBar in the decay mode $\Lambda^+_cK^-$~\cite{Aubert:2007eb} still needs more confirmations. $\Xi_c(2980)^{0,+}$ was first reported by Belle in the channels $\Lambda_c^+K^-\pi^+$ and $\Lambda_c^+K_S^0\pi^-$~\cite{Chistov:2006zj}, and was later confirmed by Belle~\cite{Yelton:2016fqw,Lesiak:2008wz} and BaBar~\cite{Aubert:2007dt} in the channels $\Xi^\prime_c(2580)\pi$, $\Xi_c(2645)\pi$ and $\Sigma_c(2455)K$, respectively. However, the decay widths reported by Refs.~\cite{Yelton:2016fqw,Chistov:2006zj,Lesiak:2008wz,Aubert:2007dt} were quite different from each other. More experimental information about the charmed baryons can be found in the review articles~\cite{Klempt:2009pi,Crede:2013sze,Amhis:2014hma,Cheng:2015iom}.

Obviously, a systematic study of masses and decays is required for these unestablished charmed baryons. More importantly, most of $2S$ and $1P$ charmed baryons have not yet been detected by any experiments. Such a research can also help the future experiments find them. In the present work, we will focus on both the mass spectra and strong decays of low-lying 1\emph{P} and 2\emph{S} charmed baryons. We pay attention to only the charmed baryons inside of which degrees of freedom of two light quarks are frozen. It means that two light quarks are not considered to be excited, neither radially nor orbitally. As illustrated in Ref.~\cite{Copley:1979wj}, this kind of charmed excitations carry lower excited energies, which means these excited charmed baryons may more likely be detected by experiments. Fortunately, our results indicate that most of the observed charmed baryons can be accommodated in this way.


The paper is arranged as follows. In Sec. \ref{sec2}, the masses of low-excited charmed baryons are calculated by the nonrelativistic quark potential model. In Sec. \ref{sec3}, the Eichten-Hill-Quigg (EHQ) decay formula which is employed to study the strong decays of excited charmed baryons is introduced. The properties of low-lying charmed baryon states are fully discussed in Sec. \ref{sec4}. Finally, the paper ends with the conclusion and outlook. Some detailed calculations and definitions are collected in Appendixes.

\section{The deduction of mass spectra}\label{sec2}

\subsection{Treating charmed baryon system as a two-body problem}

To study the baryon dynamics, one crucial question which should be answered is ``What are the relevant degrees of freedom in a baryon?''~\cite{Pervin:2007wa}. In constituent quark models, a baryon system consists of three confined quarks. Thus the dynamics of a baryon resonance is surely more complex than a meson. Due to the HQS, however, the dynamics of charmed baryons could be greatly simplified. The HQS suggests that the couplings between a \emph{c} quark and two light quarks are weak~\cite{Isgur:1991wq}. Therefore, two light quarks in a charmed baryon could first couple with each other to form a light quark cluster.\footnote{In some works, the light quark cluster may also named as a light diquark.} Then the light quark cluster couples with a charm quark, and a charmed baryon resonance forms. With this assumption, two light quarks have the same status to a \emph{c} quark, including the average distances to a \emph{c} quark. In the light cluster-heavy quark picture, the dynamics of heavy baryon can be simplified. In the nonrelativistic constituent quark model, the spin-independent parts of the Hamiltonian is
\begin{eqnarray}
H = \sum\limits_{i=1}^{3}\left(\frac{p_i^2}{2m_i}+m_i\right) + \sum\limits_{i<j}\left(-\frac{2\alpha}{3r_{ij}}+\frac{b}{2}r_{ij}\right), \label{eq1}
\end{eqnarray}
where the Cornell potential~\cite{Eichten:1978tg} is used as a phenomenological confining term. The Jacobi coordinates are usually taken to deal with the three-body problem for baryons. $\rho$, $\lambda$ and $R$ are related to quark positions by
\begin{equation}
\begin{split}
&\vec{\rho} = \vec{r}_1-\vec{r}_2,~~~~\\&
\vec{\lambda} = \frac{m_1\vec{r}_1+m_2\vec{r}_2}{m_1+m_2}-\vec{r}_Q,~~~~\\&
\vec{R} = \frac{m_1\vec{r}_1+m_2\vec{r}_2+m_Q\vec{r}_Q}{m_1+m_2+m_Q}  ,\nonumber
\end{split}
\end{equation}
where indices, 1, 2, and $Q$, are for two light quarks and a heavy quark, respectively. The momenta $\vec{p}_\rho$,  $\vec{p}_\lambda$ and $\vec{p}_R$ which are conjugate to the Jacobi coordinates above can be defined easily. Now the spin-independent Hamiltonian becomes
\begin{equation}
\begin{split}
H = &\frac{p^2_\rho}{2m_\rho} + \frac{p^2_\lambda}{2m_\lambda} + \frac{p^2_R}{2M} + M + \left(-\frac{2\alpha}{3r_{12}}+\frac{b}{2}r_{12}\right) \\&
+ \left(-\frac{2\alpha}{3r_{1Q}}+\frac{b}{2}r_{1Q}\right) + \left(-\frac{2\alpha}{3r_{2Q}}+\frac{b}{2}r_{2Q}\right), \label{eq2}
\end{split}
\end{equation}
where $m_\rho=m_1m_2/(m_1+m_2)$, $m_\lambda=(m_1+m_2)m_Q/M$, and $M=m_1+m_2+m_Q$. According to the definitions above, the relative motion between two light quarks is usually called $\rho$-mode while the one between the center of mass of the two light quarks and the heavy quark is called $\lambda$-mode.

\begin{figure}[htpb]
\begin{center}
\includegraphics[width=4.8cm,keepaspectratio]{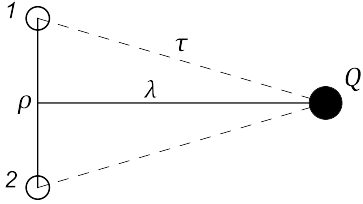}
\caption{A single heavy baryon in the light cluster-heavy quark picture. With the SU(3) flavor symmetry, the relation of $\tau=\sqrt{\lambda^2+\rho^2/4}$ is obvious.}\label{Fig1}
\end{center}
\end{figure}

As emphasized above, the average distances to \emph{c} quark should be equal for two light quarks in a cluster, \emph{i.e.}, $\tau \equiv r_{1Q}=r_{2Q}$ (see Fig.~\ref{Fig1}). In practice, we should solve the following Schr\"odinger equation for the mass of a heavy baryon,
\begin{equation}
\left[-\frac{\nabla_\rho^2}{2m_\rho}-\frac{\nabla_\lambda^2}{2m_\lambda}+\left(\frac{b\rho}{2}-\frac{2\alpha}{3\rho}\right)+\left(b\tau-\frac{4\alpha}{3\tau}\right)+C\right]\psi = E\psi. \label{eq3}
\end{equation}
Since the excited mode between two light quarks is not considered in this paper, the light quark cluster can be treated as a block with the antitriplet color structure and peculiar size. Specifically, a color singlet baryon system should be formed as $3_{q_1}\otimes3_{q_2}\otimes3_Q\ni\bar{3}_{cl.}\otimes3_Q\ni\bar{1}_{Q-cl.}$. In this way, a heavy baryon could be treated as a quasi two-body system. In the light cluster-heavy quark picture, Isgur has discussed the similarity of dynamics between heavy baryons and heavy-light mesons~\cite{Isgur:1999rf}. It should be stressed that the scenario of a light cluster-heavy quark picture is not contradictory to the one-gluon exchange interaction. Since the color-spin interaction is proportional to the inverse of quark masses, two light quarks in the heavy-baryon system are expected to strongly couple to each other~\cite{Jaffe:2004ph}. Thus, they may develop into a quark cluster. In fact, the existence of a light quark cluster correlations was partly confirmed by the lattice QCD~\cite{Alexandrou:2006cq} and the Bethe-Salpeter equation~\cite{Jinno:2015sea}.

Thus, we expect the light quark cluster to be an effective degree of freedom for a charmed baryon. Since the $\rho$ mode of a cluster is not considered here, the Schr\"odinger equation~(\ref{eq3}) is simplified as
\begin{equation}
\left(-\frac{\nabla_\lambda^2}{2m_\lambda}-\frac{4\alpha}{3\tau}+b\tau+C \right)\psi = E\psi. \label{eq4}
\end{equation}

\subsection{Adopted effective potentials}

As a whole, the light quark cluster which occupies an antitriplet color structure interacts with the $c$ quark. Then we would like to substitute $\lambda$ (distance between light cluster and $c$ quark) for $\tau$ (distance between light quark and $c$ quark). To this end, the following effective potential~\cite{Rai:2002hi}
\begin{equation}
H_{Q-cl.}^{conf}(\lambda)=-\frac{4}{3}\frac{\alpha_s}{\lambda}+b\lambda^\nu-C_{Qqq'} \label{eq5}
\end{equation}
describes the interaction between the cluster and $c$ quark, where $\nu$ is an adjustable parameter. This approximation can greatly decrease computational complexity. As shown in Tables \ref{table2} and \ref{table3}, the mass spectra given in this way are reasonable for the low-lying excited charmed baryons.

As a two-body problem, we treat the masses of different kind of clusters as parameters and first fix them before calculating the masses of low-lying charmed baryons. According to the flavor and spin, the light cluster can be classified into two kinds: one is the ``scalar" cluster, and another is the ``vector" cluster. Constrained by the Pauli's exclusion principle, the total wave function of the light quark cluster should be antisymmetric in exchange of two quarks. Because the spatial and color parts of this light quark cluster are always symmetric and antisymmetric, respectively, the function, $|\rm {flavor}\rangle\times|\rm {spin}\rangle$, should be symmetric. Therefore, the scalar light quark cluster $[qq]~(S =0)$ is always flavor antisymmetric, and the axial-vector light quark cluster $\{qq\}~(S =1)$ is flavor symmetric. In terms of the Jaffe's terminology~\cite{Jaffe:2004ph}, the ``scalar" and ``vector" quark clusters are named as the ``good" and ``bad" quark clusters, respectively. The masses of the ``good" light quark cluster $[qq]$ and $[qs]$ are taken from our previous work where $m_{[qq]}$ and $m_{[qs]}$ were fixed as 450 MeV and 630 MeV by the RFT model~\cite{Chen:2014nyo}, respectively. Following the Jaffe's method~\cite{Jaffe:2004ph}, the bad light quark cluster masses can be evaluated by the following relationships
\begin{equation}
\begin{split}
&\frac{4\times\Sigma_c(2520)+2\times\Sigma_c(2455)}{6}-\Lambda_c(2286)\approx 210~\rm {MeV},\\ &\frac{4\times\Xi^*_c(2645)+2\times\Xi'_c(2580)}{6}-\Xi_c(2470)\approx 150~\rm {MeV}. \nonumber
\end{split}
\end{equation}
Evidently, masses of $\{qq\}$ and $\{qs\}$ are about 660 MeV and 780 MeV, respectively. Henceforth, we will call the $\Lambda_c^+$ and $\Xi_c^{0,+}$ baryons the $\mathcal{G}$-type baryons, and $\Sigma_c^{0,+,++}$ and $\Xi_c^{\prime0,+}$ the $\mathcal{B}$-type baryons for convenience.

Due to an antitriplet color structure, the spin-dependent interactions between light cluster and \emph{c} quark are expected to be the same as the meson systems. In a constituent quark model~\cite{Godfrey:1985xj}, the spin-dependent interactions is written as
\begin{eqnarray}
H_S = H_{Q-cl.}^{cont}+H_{Q-cl.}^{ten}+H_{Q-cl.}^{SO}. \label{eq6}
\end{eqnarray}
The color contact interaction $H_{Q-cl.}^{cont}$ is usually given by the following form
\begin{equation}
H_{Q-cl.}^{cont}=\frac{32\pi}{9}\frac{\alpha_s}{m_Qm_{cl.}}\tilde{\delta}_{\sigma}(\lambda)\vec{S}_Q\cdot\vec{S}_{cl.}, \label{eq7}
\end{equation}
where $\vec{S}_Q$ and $\vec{S}_{cl.}$ refer to the heavy quark and light cluster spins. A Gaussian-smeared function $({\sigma}/\sqrt{\pi})^3e^{-{\sigma}^2\lambda^2}$ is normally used for $\tilde{\delta}_{\sigma}(\lambda)$~\cite{Barnes:2005pb}. If the SU(3) flavor symmetry is kept well for the
charmed baryons,
we may modify the color contact interaction as
\begin{equation}
H_{Q-cl.}^{cont}=\frac{32}{9\sqrt{\pi}}\frac{\alpha_s\sigma^3}{m_Q}e^{-\sigma^2\lambda^2}\vec{S}_Q\cdot\vec{S}_{cl.}, \label{eq8}
\end{equation}
where the mass of a light quark cluster is just replaced by a unit. This assumption is supported by the mass differences of the $1S$ $\mathcal{B}-$type 
charmed baryons,
\begin{equation}
\begin{split}
&\Sigma_c(2520)^{++}-\Sigma_c(2455)^{++}=64.44^{+0.25}_{-0.24}~\rm {MeV},~~~~\\&
\Xi_c(2645)^{+}~-\Xi'_c(2580)^{+}~~=70.2\pm3.0~\rm {MeV},~~~~\\&
\Omega_c(2770)^0~-\Omega_c(2695)^0~~=70.7\pm2.6~\rm {MeV}  .\nonumber
\end{split}
\end{equation}
The mass differences shown above are mainly due to the color contact interaction in the quark potential model. Clearly, these values are almost independent of the light quark cluster masses. The color tensor interaction in Eq.~(\ref{eq6}) is
\begin{equation}
H_{Q-cl.}^{ten}=\frac{4}{3}\frac{\alpha_s}{m_Qm_{cl.}}\frac{1}{\lambda^3}\left(\frac{3\left(\vec{S}_Q\cdot\vec{\lambda}\right)\left(\vec{S}_{cl.}\cdot\vec{\lambda}\right)}{\lambda^2}-\vec{S}_Q\cdot\vec{S}_{cl.}\right), \label{eq9}
\end{equation}
Finally, $H_{Q-cl.}^{SO}$ denotes the spin-orbit interaction which contains two terms. One is the color magnetic interaction which arises from one-gluon exchange
\begin{equation}
H_{Q-cl.}^{SO(cm)}=\frac{4}{3}\frac{\alpha_s}{\lambda^3}\left(\frac{\vec{S}\cdot\vec{L}}{m_Qm_{cl.}}+\frac{\vec{S}_Q\cdot\vec{L}}{m_Q^2}+\frac{\vec{S}_{cl.}\cdot\vec{L}}{m_{cl.}^2}\right), \label{eq10}
\end{equation}
where $\vec{S}$ denotes the spin of a baryon, $\vec{S}=\vec{S}_Q+\vec{S}_{cl.}$. Another spin-orbit interaction is the Thomas-precession term
\begin{equation}
H_{Q-cl.}^{SO(tp)}=-\frac{1}{2\lambda}\frac{\partial H_{Q-cl.}^{conf}}{\partial \lambda}\left(\frac{\vec{S}_Q\cdot\vec{L}}{m_Q^2}+\frac{\vec{S}_{cl.}\cdot\vec{L}}{m_{cl.}^2}\right). \label{eq11}
\end{equation}

To reflect the importance of the heavy quark symmetry, we rewrite the spin-dependent interactions as
\begin{equation}
H_S=V_{ss}~\vec{S}_Q\cdot\vec{S}_{cl.}+V_1~\vec{S}_{cl.}\cdot\vec{L}+V_2~\vec{S}_Q\cdot\vec{j}_{cl.}+V_t~\hat{S}_{12}. \label{eq12}
\end{equation}
The degrees of freedom of the light quark cluster are characterized by its total angular momentum $\vec{j}_{cl.}$, \emph{i.e.}, $\vec{j}_{cl.}=\vec{S}_{cl.}+\vec{L}$. Obviously, the orbital angular momentum $\vec{L}$ of a charmed baryon in the present picture is defined by the angular momentum between light quark cluster and \emph{c} quark, \emph{i.e.}, $\vec L=\vec L_\lambda$. The tensor operator is defined as $\hat{S}_{12}=3\left(\vec{S}_Q\cdot\vec{\lambda}\right)\left(\vec{S}_{cl.}\cdot\vec{\lambda}\right)/\lambda^2-\vec{S}_Q\cdot\vec{S}_{cl.}$.

With the confining term of Eq. (\ref{eq5}), the coefficients $V_{ss}$, $V_1$, $V_2$, and $V_t$ in Eq. (\ref{eq12}) are defined by
\begin{equation}
\begin{split}
&V_{ss}=\frac{1}{m_Q}\left[\frac{32\alpha_s}{9\sqrt{\pi}}\sigma^3e^{-\sigma^2\lambda^2}-\frac{1}{m_Q}\left(\frac{2\alpha_s}{3\lambda^3}-\frac{b\nu}{2}\lambda^{\nu-2}\right)-\frac{4\alpha_s}{3\lambda^3}\frac{1}{m_{cl.}}\right],~~~~\\&
V_1=\frac{1}{m_{cl.}}\left[\frac{1}{m_{cl.}}\left(\frac{2\alpha_s}{3\lambda^3}-\frac{b\nu}{2}\lambda^{\nu-2}\right)+\frac{4\alpha_s}{3\lambda^3}\frac{1}{m_{Q}}\right],~~~~\\&
V_2=\frac{1}{m_Q}\left[\frac{1}{m_Q}\left(\frac{2\alpha_s}{3\lambda^3}-\frac{b\nu}{2}\lambda^{\nu-2}\right)+\frac{4\alpha_s}{3\lambda^3}\frac{1}{m_{cl.}}\right],~~~~\\&
V_t=\frac{4\alpha_s}{3\lambda^3}\frac{1}{m_Qm_{cl.}}.  \label{eq13}
\end{split}
\end{equation}

\subsection{Getting masses of charmed baryons}

In our calculation, the following Schr\"odinger equation is solved for the $nS$ state:
\begin{equation}
\left(-\frac{\nabla_\lambda^2}{2m_\lambda}+H_{Q-cl.}^{conf}+H_{Q-cl.}^{cont}\right)\Psi=E\Psi. \label{eq14}
\end{equation}
The confining and contact terms have been given by Eqs.~(\ref{eq5}) and~(\ref{eq8}). For the orbital excitations, all spin-dependent interactions are treated as the leading-order perturbations. Our calculation indicates that the color contact interaction can be ignored for the orbital excitations.

Two bases are employed to extract the mass matrix elements. One is the eigenstates $|S_{cl.}, L, j_{cl.}, S_Q, J\rangle$ ($jj$ coupling scheme) and another is $|S_{cl.}, S_Q, S, L, J\rangle$  ($LS$ coupling scheme). The relation between these two bases is
\begin{equation}
\begin{split}
|[S_{cl.}, L]_{j_{cl.}}, S_Q\rangle_J=\sum\limits_S&(-1)^{S_{cl.}+S_Q+L+J}\sqrt{(2J_{cl.}+1)(2S+1)}\\ &\times\left\{
           \begin{array}{ccc}
                    s_{cl.}  & L    & j_{cl.}\\
                    J       & S_Q    & S\\
                    \end{array}
     \right\}|[S_{cl.}, S_Q]_S, L\rangle_J,\label{eq15}
\end{split}
\end{equation}

\begin{figure}[htpb]
\begin{center}
\includegraphics[width=8.4cm,keepaspectratio]{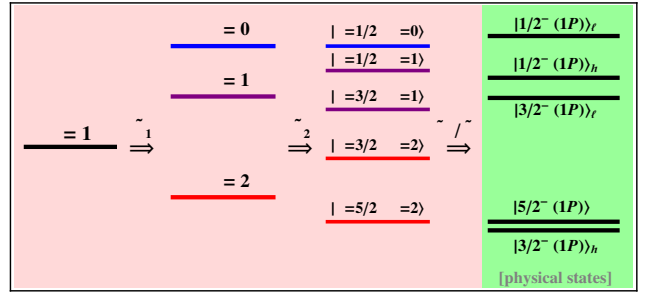}
\caption{A schematic diagram for the splittings of \emph{p}-wave $\Sigma_c$ and $\Xi'_c$. Here $\vec j_l=\vec L+\vec j_{cl.}$ and subindices $\ell$ and $h$ of the last column mean low and high states in mass after including $\tilde{V}_{ss}$ and $\tilde{V}_t$ interactions.}\label{Fig2}
\end{center}
\end{figure}

Due to $S_{cl.}=0$, only $V_2~\vec{S}_Q\cdot\vec{j}_{cl.}$ contributes to the masses of $\mathcal{G}$-type charmed baryons. With a bad light quark cluster, however, $\mathcal{B}$-type charmed baryons have more complicated splitting structures. Within the framework of the heavy quark effective theory, the spin of an axial-vector light quark cluster, $S_{cl.}$, first couples with the orbital angular momentum $L$. As illustrated in Fig. \ref{Fig2}, in the heavy quark limit $m_c\rightarrow\infty$, there are only three states which are characterized by $j_{cl.}$ for $1P$ charmed baryons. When the heavy quark spin $S_Q$ couples with $j_{cl.}$, the degeneracy is resolved and the five states appear. They are two $J^P=1/2^-$, two $J^P=3/2^-$, and one $J^P=5/2^-$ states. Lastly, the states with the same $J^P$ mix with each other by the interactions of $V_{ss}~\vec{S}_Q\cdot\vec{S}_{cl.}$ and $V_t~\hat{S}_{12}$, and physical states are formed.

We now turn to a calculation of the mass matrix in the $jj$ coupling scheme. For 1\emph{P} states with $J^P=1/2^-$, the mass matrix is given by
\begin{eqnarray}
\langle\Phi_{1/2}\mid H_S\mid\Phi_{1/2}\rangle=\left(
           \begin{array}{ccc}
                    -2V_1-4V_t  & \frac{V_{ss}-4V_t}{\sqrt{2}}\\
                    \frac{V_{ss}-4V_t}{\sqrt{2}}  & -V_1-V_2-\frac{V_{ss}}{2}-2V_t \\
                    \end{array}
     \right).\nonumber
\end{eqnarray}
Similarly, for two states with $J^P=3/2^-$,
\begin{eqnarray}
\begin{split}
\langle\Phi_{3/2} &\mid H_S\mid\Phi_{3/2}\rangle\\ &=\left(
           \begin{array}{ccc}
                    -V_1+\frac{V_2}{2}+\frac{V_{ss}}{4}+4V_t  & \frac{5V_{ss}+16V_t}{4\sqrt{5}}\\
                    \frac{5V_{ss}+16V_t}{4\sqrt{5}}  & V_1-\frac{3V_2}{2}-\frac{3V_{ss}}{4}+\frac{4V_t}{5} \\
                    \end{array}
     \right).\nonumber
\end{split}
\end{eqnarray}
For the $J^P=5/2^-$ state,
\begin{equation}
\begin{split}
\langle j_l=2, J^P=5/2^- &\mid H_S\mid j_l=2,  J^P=5/2^-\rangle\\ &=V_1+V_2+\frac{1}{2}V_{ss}-\frac{6}{5}V_{t}.\nonumber
\end{split}
\end{equation}
In the following, we denote $|S_{cl.}, L, j_{cl.}, S_Q, J\rangle$ as $|j_{cl.}, J^P\rangle$ for brevity. Then the notations $\mid\Phi_{1/2}\rangle$ and $\mid\Phi_{3/2}\rangle$ appearing above are defined by
\begin{eqnarray}
\begin{split}
\mid\Phi_J\rangle
=\left(
           \begin{array}{cc}
                    \mid j_l=J-1/2, J^P\rangle  \\
                    \mid j_l=J+1/2, J^P\rangle  \\
                    \end{array}
     \right). \nonumber
\end{split}
\end{eqnarray}
The mass matrix of 1\emph{D} states can also be obtained by the similar procedure. As shown above, there are seven parameters in the nonrelativistic quark potential model, which are $m_Q$, $m_{cl.}$, \emph{b}, $\alpha$, $\gamma$, $\nu$, and $C_{Qqq'}$. All values of parameters are listed in Table~\ref{table1}. If the SU(3) flavor symmetry is taken into account for the charmed and charmed-strange baryons, the dynamics of $\Lambda_c^+$ states should be like $\Xi_c$. The case of $\Sigma_c$ and $\Xi'_c$ is alike. Accordingly, the same value of $\gamma$ is selected for the $\mathcal{G}$-type charmed baryons, as well as the case of $\mathcal{B}$-type.
 \begin{table}[htbp]
\caption{Values of the parameters of the nonrelativistic quark potential model. The unit of \emph{b} is GeV$^{\nu+1}$ which varies depending on each value of $\nu$.
}\label{table1}
\renewcommand\arraystretch{1.2}
\begin{tabular*}{82mm}{l@{\extracolsep{\fill}}cllcc}
\toprule[1pt]\toprule[1pt]
$m_c$             & 1.68~GeV              & \emph{b}                            & 0.145            & $C_{\Lambda_C}$   & 0.233~GeV\\
$m_{[qq]}$        & 0.45~GeV              & $\alpha$                            & 0.45             & $C_{\Sigma_C}$    & 0.100~GeV\\
$m_{[qs]}$        & 0.63~GeV              & $\nu_{[\Lambda_c,\Xi_c]}$           & 0.84             & $C_{\Xi_C}$       & 0.156~GeV\\
$m_{\{qq\}}$      & 0.66~GeV              & $\nu_{[\Sigma_c,\Xi'_c]}$           & 0.70             & $C_{\Xi'_C}$      & 0.060~GeV\\
$m_{\{qs\}}$      & 0.78~GeV              & $\sigma$                            & 1.00~GeV         &                   &  \\
\bottomrule[1pt]\bottomrule[1pt]
\end{tabular*}
\end{table}

We have adopted the typical values in the quark potential models for $m_c$, \emph{b}, $\alpha$, and $\nu$ (see Table \ref{table1}). It is an effective method to investigate charmed baryons in heavy quark-light quark cluster picture. We do not expect the values of $\nu$ to be the same both for $\mathcal{G}$-type and $\mathcal{B}$-type baryons. Here, $\nu$ of $\Lambda_c^+/\Xi_c$ is slightly lager than $\Sigma_c/\Xi'_c$. The predicted masses of low excited charmed baryons are collected in Tables \ref{table2} and \ref{table3}.

\begin{table*}[htbp]
\caption{Predicted masses for $\Lambda_c^+$ and $\Xi_c$ states of ours and other approaches in Refs.~\cite{Ebert:2011kk,Chen:2014nyo,Capstick:1986bm,Roberts:2007ni} compared to experimental data~\cite{Agashe:2014kda} (in MeV).} \label{table2}
\renewcommand\arraystretch{1.2}
\begin{tabular*}{170mm}{@{\extracolsep{\fill}}lcccccccccc}
\toprule[1pt]\toprule[1pt]
\multirow{2}{*}{States}  & \multicolumn{5}{c}{$\Lambda_c^+$ baryons}  & \multicolumn{5}{c}{$\Xi_c$ baryons}  \\
\cline{2-6}\cline{7-11}
    & PDG~\cite{Agashe:2014kda}  &   Prediction  &Ref.~\cite{Ebert:2011kk}& Ref.~\cite{Chen:2014nyo}& Ref.~\cite{Capstick:1986bm} & PDG~\cite{Agashe:2014kda} &   Prediction  &Ref.~\cite{Ebert:2011kk}& Ref.~\cite{Chen:2014nyo}& Ref.~\cite{Roberts:2007ni}  \\
\cline{1-6}\cline{7-11}
$\mid 1S, 1/2^+\rangle$ &  2286.86  &  2286      &  2286   &  2286   &  2265  &  2470.88 &  2470  &  2476     &  2467     &  2466    \\
$\mid 2S, 1/2^+\rangle$ &  2766.6   &  2772      &  2769   &  2766   &  2775  &  2968.0  &  2940  &  2959     &  2959     &  2924    \\
$\mid 3S, 1/2^+\rangle$ &           &  3116      &  3130   &  3112   &  3170  &          &  3265  &  3323     &  3325     &         \\
$\mid 1P, 1/2^-\rangle$ &  2592.3   &  2614      &  2598   &  2591   &  2630  &  2791.8  &  2793  &  2792     &  2779     &  2773   \\
$\mid 1P, 3/2^-\rangle$ &  2628.1   &  2639      &  2627   &  2629   &  2640  &  2819.6  &  2820  &  2819     &  2814     &  2783    \\
$\mid 1D, 3/2^+\rangle$ &           &  2843      &  2874   &  2857   &  2910  &  3054.2  &  3033  &  3059     &  3055     &  3012    \\
$\mid 1D, 5/2^+\rangle$ &  2881.53  &  2851      &  2880   &  2879   &  2910  &  3079.9  &  3040  &  3076     &  3076     &  3004   \\
$\mid 2P, 1/2^-\rangle$ &  2939.3   &  2980      &  2983   &  2989   &  3030  &  3122.9  &  3140  &  3179     &  3195     &         \\
$\mid 2P, 3/2^-\rangle$ &           &  3004      &  3005   &  3000   &  3035  &          &  3164  &  3201     &  3204     &         \\
\bottomrule[1pt]\bottomrule[1pt]
\end{tabular*}
\end{table*}

\begin{table*}[htbp]
\caption{Predicted masses for $\Sigma_c$ and $\Xi'_c$ states of ours and other approaches in Refs.~\cite{Ebert:2007nw,Ebert:2011kk,Capstick:1986bm,Shah:2016mig} compared to experimental data~\cite{Agashe:2014kda} (in MeV).} \label{table3}
\renewcommand\arraystretch{1.2}
\begin{threeparttable}
\begin{tabular*}{170mm}{@{\extracolsep{\fill}}lcccccccccc}
\toprule[1pt]\toprule[1pt]
\multirow{2}{*}{States}  & \multicolumn{6}{c}{$\Sigma_c$ baryons}  & \multicolumn{4}{c}{$\Xi'_c$ baryons}  \\
\cline{2-7}\cline{8-11}
    & PDG~\cite{Agashe:2014kda}  &   Prediction  & Ref.~\cite{Ebert:2007nw} &Ref.~\cite{Ebert:2011kk} &Ref.~\cite{Capstick:1986bm}& Ref.~\cite{Shah:2016mig} & PDG~\cite{Agashe:2014kda} &   Prediction  & Ref.~\cite{Ebert:2007nw} &Ref.~\cite{Ebert:2011kk}   \\
\cline{1-7}\cline{8-11}
$\mid 1S, 1/2^+\rangle$        &  2452.9   &  2456      &  2439   &  2443   &  2440  &  2452  &  2575.6  &  2579  &  2579     &  2579          \\
$\mid 1S, 3/2^+\rangle$        &  2517.5   &  2515      &  2518   &  2519   &  2495  &  2501  &  2645.9  &  2649  &  2654     &  2649         \\
$\mid 2S, 1/2^+\rangle$        &  2846$^a$ &  2850      &  2864   &  2901   &  2890  &  2961  &          &  2977  &  2984     &  2983         \\
$\mid 2S, 3/2^+\rangle$        &           &  2876      &  2912   &  2936   &  2985  &  2996  &          &  3007  &  3035     &  3026          \\
$\mid 3S, 1/2^+\rangle$        &           &  3091      &         &  3271   &  3035  &  3381  &          &  3215  &           &  3323        \\
$\mid 3S, 3/2^+\rangle$        &           &  3109      &         &  3293   &  3200  &  3403  &          &  3236  &           &  3396         \\
$\mid 1P, 1/2^-\rangle_l$      &           &  2702      &  2795   &  2713   &  2765  &  2832  &          &  2839  &  2928     &  2854        \\
$\mid 1P, 1/2^-\rangle_h$      &  2766.6   &  2765      &  2805   &  2799   &  2770  &  2841  &          &  2900  &  2934     &  2936          \\
$\mid 1P, 3/2^-\rangle_l$      &           &  2785      &  2761   &  2773   &  2770  &  2812  &  2931    &  2921  &  2900     &  2912         \\
$\mid 1P, 3/2^-\rangle_h$      &  2801     &  2798      &  2798   &  2798   &  2805  &  2822  &          &  2932  &  2931     &  2935         \\
$\mid 1P, 5/2^-\rangle$        &           &  2790      &  2799   &  2789   &  2815  &  2796  &          &  2927  &  2921     &  2929          \\
$\mid 1D, 1/2^+\rangle$        &           &  2949      &  3014   &  3041   &  3005  &        &          &  3075  &  3132     &  3163          \\
$\mid 1D, 3/2^+\rangle_l$      &           &  2952      &  3005   &  3040   &  3060  &        &          &  3089  &  3127     &  3160         \\
$\mid 1D, 3/2^+\rangle_h$      &           &  2964      &  3010   &  3043   &  3065  &        &          &  3081  &  3131     &  3167          \\
$\mid 1D, 5/2^+\rangle_l$      &           &  2942      &  2960   &  3023   &  3065  &        &          &  3091  &  3087     &  3153        \\
$\mid 1D, 5/2^+\rangle_h$      &           &  2962      &  3001   &  3038   &  3080  &        &          &  3077  &  3123     &  3166         \\
$\mid 1D, 7/2^+\rangle$        &           &  2943      &  3015   &  3013   &  3090  &        &          &  3078  &  3136     &  3147         \\
$\mid 2P, 1/2^-\rangle_l$      &           &  2971      &  3176   &  3125   &  3185  &  3245  &          &  3094  &  3294     &  3267             \\
$\mid 2P, 1/2^-\rangle_h$      &           &  3018      &  3186   &  3172   &  3195  &  3256  &          &  3144  &  3300     &  3313               \\
$\mid 2P, 3/2^-\rangle_l$      &           &  3036      &  3147   &  3151   &  3195  &  3223  &          &  3172  &  3269     &  3293              \\
$\mid 2P, 3/2^-\rangle_h$      &           &  3044      &  3180   &  3172   &  3210  &  3233  &          &  3165  &  3296     &  3311              \\
$\mid 2P, 5/2^-\rangle$        &           &  3040      &  3167   &  3161   &  3220  &  3203  &          &  3170  &  3282     &  3303              \\
\bottomrule[1pt]\bottomrule[1pt]
\end{tabular*}
\begin{tablenotes}
        \footnotesize
        \item[a] The mass value for the $\mid 2S, 1/2^+\rangle$ state is taken from the measurement of BaBar~\cite{Aubert:2008ax}.
      \end{tablenotes}
\end{threeparttable}
\end{table*}

 As mentioned earlier, the nonzero off-diagonal elements in mass matrices of $\langle\Phi_{1/2}\mid H_S\mid\Phi_{1/2}\rangle$ and $\langle\Phi_{3/2}\mid H_S\mid\Phi_{3/2}\rangle$ cause the mixing between two states with the same $J^P$ but different $j_{cl.}$. However, the mechanism of mixing effects in hadron physics is still unclear. In principle, a physical hadron state with a specific $J^P$ comprises all possible Fock states with the same total spin and parity. As the most famous member of the \emph{XYZ} family, $X(3872)$ may be explained as a mixture between charmonium and molecular state with $J^{PC}=1^{++}$~\cite{Matheus:2009vq}. Here we take the $|j_{cl.}, J^P\rangle$ basis to describe the mixing for the $\mathcal{B}-$type baryons. Then two physical states characterized by different masses can be denoted as
\begin{eqnarray}
\begin{aligned}
 \left(
           \begin{array}{c}
                     |High, J^P\rangle\\
                     |Low, J^P\rangle\\
                    \end{array}
     \right)&=\left(
           \begin{array}{cc}
                    ~\cos\phi  & \sin\phi \\
                    -\sin\phi  & \cos\phi\\
                    \end{array}
     \right)  \left(
           \begin{array}{c}
                     |J-1/2, J^P\rangle \\
                     |J+1/2, J^P\rangle \\
                    \end{array}
     \right). \label{eq16}
\end{aligned}
\end{eqnarray}
For example, two 1\emph{P} $\Sigma_c$ states with $J^P=1/2^-$ can be represented as
\begin{eqnarray}
\begin{aligned}
 \left(
           \begin{array}{c}
                     \Sigma_c(2765)\\
                     \Sigma_c(2702)\\
                    \end{array}
     \right)&=\left(
           \begin{array}{cc}
                    ~\cos125.4^\circ  & \sin125.4^\circ \\
                    -\sin125.4^\circ  & \cos125.4^\circ\\
                    \end{array}
     \right)  \left(
           \begin{array}{c}
                     |0, 1/2^-\rangle \\
                     |1, 1/2^-\rangle \\
                    \end{array}
     \right). \label{eq17}
\end{aligned}
\end{eqnarray}
Here we have denoted the physical states by their masses (see Table \ref{table3}). The mixing angles for other states in Table \ref{table3} with the same $J^P$ are listed in Table~\ref{table4}.

\begin{table}[htbp]
\caption{The mixing angles for the $1P$, $2P$, and $1D$ $\Sigma_c/\Xi'_c$ states.
}\label{table4}
\renewcommand\arraystretch{1.2}
\begin{tabular*}{85mm}{c@{\extracolsep{\fill}}cccccc}
\toprule[1pt]\toprule[1pt]
                 &   $1P(1/2^-)$  &   $1P(3/2^-)$  &   $2P(1/2^-)$  &   $2P(3/2^-)$  &   $1D(3/2^+)$  &   $1D(5/2^+)$   \\
\midrule[0.8pt]
$\Sigma_c$      & $125.4^\circ$  & $-156.8^\circ$  & $124.8^\circ$  & $-151.4^\circ$  & $172.2^\circ$  & $-175.6^\circ$  \\
$\Xi'_c$         & $125.0^\circ$  & $-153.6^\circ$  & $124.3^\circ$  & $-145.1^\circ$  & $168.9^\circ$  & $-173.8^\circ$  \\
\bottomrule[1pt]\bottomrule[1pt]
\end{tabular*}
\end{table}

Our results of mixing angles in Table~\ref{table4} indicate that the heavier $1/2^-$ state, $\Sigma_c(2765)$, is dominated by a $|1, 1/2^-\rangle$ component, while $\Sigma_c(2702)$ is by a $|0, 1/2^-\rangle$ component. For two $3/2^-$ states, the light $\Sigma_c(2785)$ is dominated by $|2, 1/2^-\rangle$, while the heavy $\Sigma_c(2798)$ by $|1, 1/2^-\rangle$. The mixing of $2P$ states is similar to the $1P$ states. For the $1D$ states, one notices that both $3/2^+$ and $5/2^+$ with heavier masses are dominated by smaller $j_{cl.}$ components.

The uncertainty may exist in the mixing angles. Firstly, the loop corrections to the spin-dependent one-gluon-exchange potential may be important for the heavy-light hadrons. As an example, the lower mass of $D_s(2317)^\pm$ compared with the old calculations~\cite{Godfrey:1985xj} can be well explained by the corrected spin-dependent potential~\cite{Lakhina:2006fy,Radford:2009bs}. If we use this type of potential in our calculation, of course, the mixing angle will change. Secondly, the mixing angles depend on the parameters. Thirdly, there are other mechanisms, \emph{e.g}., hadron loop effects~\cite{Close:2009ii}, which may contribute to the mixing phenomenon in hadron physics. Anyway, we expect that the mixing angles in Table \ref{table4} reflect main features of the mixing states. Due to the uncertainties of the mixing angles, however, we ignore the mixing effects as the first step to study the decays of charmed excitations in the next Subsection. Obviously, it is a good approximation only when the mixing effects are not large. Fortunately, this crude procedure is partially supported by the former analysis of charmed mesons~\cite{Chen:2011rr,Chen:2012zk,Chen:2015lpa}. If the decay properties obtained in this way describe principal characteristics of the mixing states, the angles obtained by the potential model may be overestimated.

\subsection{Simple harmonic oscillator (SHO) $\beta$ values}

In the next Section, the Okubo-Zweig-Iizuka (OZI) allowed decays will be calculated for the $1P$ and $2S$ charmed baryons where the SHO wave functions are used to evaluate the transition factors via the $^3P_0$ model. We will also discuss the mixing effects for the decays of the relevant states. Following the method of Ref.~\cite{Close:2005se}, all values of the SHO wave function scale, denoted as $\beta$ in the following, are calculated (see Table \ref{table5}). The values of $\beta$ reflect the distances between the light quark cluster and \emph{c} quark.
\begin{table}[htbp]
\caption{The meson effective $\beta$ values in GeV.
}\label{table5}
\renewcommand\arraystretch{1.2}
\begin{tabular*}{82mm}{c@{\extracolsep{\fill}}ccccc}
\toprule[1pt]\toprule[1pt]
\multicolumn{2}{c}{States}            & $\Lambda_c^+$         &   $\Xi_c$              & $\Sigma_c$  & $\Xi'_c$      \\
\midrule[0.8pt]
\multirow{2}{*}{1S}       & $1/2^+$   & \multirow{2}{*}{0.291}& \multirow{2}{*}{0.331} & 0.335       &    0.362      \\
                          & $3/2^+$   &                       &                        & 0.296       &    0.315     \\
\multicolumn{2}{c}{2S}                &    0.145              & 0.162                  & 0.144       &    0.152      \\
\multicolumn{2}{c}{3S}                &    0.102              & 0.113                  & 0.098       &    0.103    \\
\multicolumn{2}{c}{1P}                &    0.184              & 0.205                  & 0.182       &    0.192    \\
\multicolumn{2}{c}{2P}                &    0.117              & 0.130                  & 0.112       &    0.118    \\
\multicolumn{2}{c}{1D}                &    0.142              & 0.156                  & 0.136       &    0.143    \\
\bottomrule[1pt]\bottomrule[1pt]
\end{tabular*}
\end{table}

In our calculation of strong decays, we will consider the structures of light diquarks. What is more important is that the possible final states of an excited charmed baryon may contain a light flavor meson, a charmed meson, a light flavor baryon, \emph{e.g}., $\pi$, \emph{K}, \emph{D}, \emph{p}, and $\Lambda$. For the $\beta$ of these hadrons, the following potential will be used
\begin{equation}
V(r)=\textbf{F}_{q_1} \cdot \textbf{F}_{q_2}\left(\frac{\alpha_s}{r}-\frac{3}{4}br+\frac{3}{4}C+\frac{32\alpha_s\sigma^3e^{-\sigma^2r^2}}{9\sqrt{\pi}m_qm_q}\vec{S}_{q_1} \cdot \vec{S}_{q_2}\right), \label{eq18}
\end{equation}
where
\begin{subnumcases}{\langle\textbf{F}_{q_1} \cdot \textbf{F}_{q_2}\rangle~=~}
 -\frac{4}{3} & for~~$q_1\bar{q}_2$ \\
 -\frac{2}{3} & for~~$q_1q_2$
  \label{eq19}
 \end{subnumcases}

Here, the parameters $\alpha_s$ and \emph{b} are taken as 0.45 and 0.145 GeV$^{\nu+1}$ as in Table~\ref{table1}, respectively. To reproduce the masses of light quark clusters in Table~\ref{table1}, the masses of $u/d$, $s$ are fixed as 0.195 GeV and 0.380 GeV. While $\sigma$ and \emph{C} are treated as adjustable parameters, the masses of $\pi/\rho$, $K/K^*$, $D/D^*$, $p/\Delta$, and $\Lambda$ families are fitted with experimental data. Meanwhile, the values of $\beta$ for the corresponding states are also obtained, which are collected in Table \ref{table6}.
\begin{table*}[htbp]
\caption{The effective $\beta$ values in GeV for the light quark cluster and various hadrons (the second row). The values of $\sigma$ and \emph{C} are given in the square brackets for various hadron structures (the third row). }\label{table6}
\renewcommand\arraystretch{1.5}
\begin{center}
\begin{tabular*}{166mm}{c@{\extracolsep{\fill}}ccccccccccc}
\toprule[1pt]\toprule[1pt]
 $[qq]$  & $\{qq\}$ &  $[qs]$ &  $\{qs\}$  &  $\pi$ &  $\rho$ &  $K$ &  $K^*$  &  $D$ &  $D^*$   &  $p$ &  $\Lambda$\\
 \midrule[0.7pt]
 0.201   & 0.143    &  0.207  &  0.159     &  0.298 &  0.179  & 0.291& 0.201   &0.250 &  0.230   &0.189 &  0.226\\
\multicolumn{2}{c}{[1.17,~0.39]}    &\multicolumn{2}{c}{[1.57,~0.38]}     &\multicolumn{2}{c}{[0.73,~0.63]}  &\multicolumn{2}{c}{[0.83,~0.48]}   &\multicolumn{2}{c}{[1.20,~0.63]}  &[$-$,~0.38] &  [$-$,~0.26]\\
\bottomrule[1pt]\bottomrule[1pt]
\end{tabular*}
\end{center}
\end{table*}

Before ending this section, we briefly summarize the complicated deduction presented here. Firstly, the dynamics of heavy baryon is simplified as a two-body system when the symmetric configuration is considered. Secondly, the mass matrices were calculated in the $jj$ coupling scheme. By solving the Schr\"odinger equation, we obtained the mass spectra and mixing angles for the relevant states. For estimating the two-body strong decays in next Section, finally, we also presented the values of the SHO wave function scale for all initial and final states.

\section{Strong Decays}\label{sec3}

In this section, we will use the formula provided by Eichten, Hill, and Quigg (EHQ)~\cite{Eichten:1993ub} to extract the decay widths of excited charmed baryons. Since the dynamical behavior of the heavy-light hadrons is governed by the light degrees of freedom in the limit of heavy quark symmetry, a doublet formed by two states with the same ${j}_{cl.}$ but different $J$ shall have the similar decay properties. More specifically, the transitions between two doublets should be determined by a single amplitude which is proportional to the products of four Clebsch-Gordan coefficients~\cite{Isgur:1991wq}. Some typical ratios of excited charmed baryons with negative-parity were predicted by this law~\cite{Isgur:1991wq}. Later, a more concise formula (the EHQ formula) was proposed for the widths of heavy-light mesons~\cite{Eichten:1993ub}. The EHQ formula has been applied systematically to the decays of excited open-charm mesons~\cite{Chen:2011rr,Chen:2012zk,Chen:2015lpa}. Recently, the EHQ formula has been extended to study the decay properties of 1\emph{D} $\Lambda_c$ and $\Xi_c$ states~\cite{Chen:2014nyo}.

\begin{figure}[htpb]
\begin{center}
\includegraphics[width=7.8cm,keepaspectratio]{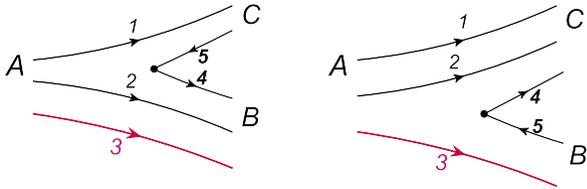}
\caption{The two topological diagrams for an excited charmed baryon
\emph{A} decaying into the final states $B$ and $C$. The brown line 3 denotes a charm quark.}\label{Fig3}
\end{center}
\end{figure}

For the charmed baryons, the EHQ formula can be written as
\begin{equation}
\Gamma^{A\rightarrow BC}_{j_C,\ell} = \xi\,\left(\mathcal
{C}^{s_Q,j_B,J_B}_{j_C,j_A,J_A}\right)^2\left|\mathcal
{M}^{j_A,j_B}_{j_C,\ell}(q)\right|^2 \,q^{2\ell+1}\,
e^{-q^2/\tilde{\beta}^2}, \label{eq20}
\end{equation}
where $\xi$ is the flavor factor given in Table \ref{table13} in Appendix \ref{2}. $q=|\vec q|$ denotes the three-momentum of a final state in the rest frame of an initial state. \emph{A} and $B$ represent the initial and final heavy-light hadrons, respectively. $C$ denotes the light flavor hadron (see Fig.\ref{Fig3}). The explicit expression of $\tilde{\beta}$ is given in Eq. (\ref{A11}) in Appendix \ref{1}. In addition, $\mathcal
{C}^{s_Q,j_B,J_B}_{j_C,j_A,J_A}$ is a normalized coefficient given by the following equation,
\begin{eqnarray}\label{eq21}
\begin{split}
\mathcal {C}^{s_Q,j_B,J_B}_{j_C,j_A,J_A}=(-1)^{J_A+j_B+j_C+s_Q}~&\sqrt{(2j_A+1)(2J_B+1)}\\ &\times\left\{
           \begin{array}{ccc}
                    s_Q  & j_B & J_B\\
                    j_C  & J_A & j_A\\
                    \end{array}
     \right\},
\end{split}
\end{eqnarray}
where $\vec{j}_C \equiv \vec{s}_C + \vec{\ell}$. The symbols $s_C$ and $\ell$ represent the spin of the light hadron $C$ and the orbital angular momentum relative to $B$, respectively. The transition factors $\mathcal {M}^{j_A,j_B}_{j_C,\ell}(q)$ involved in the concrete dynamics can only be calculated by various phenomenological models. For the decays of heavy-light mesons, transition factors have been calculated by the relativistic chiral quark model~\cite{Goity:1998jr} and the $^3P_0$ model~\cite{Page:1998pp,Chen:2011rr,Chen:2015lpa}. In our work, we will employ the $^3P_0$ model~\cite{Micu:1968mk,LeYaouanc:1972vsx,LeYaouanc:1988fx} to obtain the transition factors. More details for an estimate of the transition factors are given in Appendix~\ref{1}.

\section{Discussion}\label{sec4}

\subsection{Experimentally well established 1\emph{S} and 1\emph{P} states}\label{a}

\begin{table}[htbp]
\caption{Open-flavor strong decay widths of 1\emph{S} $\Sigma_c$ and $\Xi'_c$ in MeV.}\label{table7}
\renewcommand\arraystretch{1.2}
\begin{tabular*}{85mm}{c@{\extracolsep{\fill}}ccccccc}
\toprule[1pt]\toprule[1pt]
                 \multicolumn{8}{c}{1\emph{S} $\Sigma_c$ and $\Xi'_c$}   \\
\midrule[0.8pt]
\multicolumn{4}{c}{ $1/2^+$}  & \multicolumn{4}{c}{$3/2^+$} \\
\cline{1-4}\cline{5-8}
\multicolumn{2}{c}{$\Sigma_c(2455)^{++}$}  & \multicolumn{2}{c}{$\Xi'_c(2580)^+$}  & \multicolumn{2}{c}{$\Sigma_c(2520)^{++}$}  & \multicolumn{2}{c}{$\Xi'_c(2645)^+$} \\
\cline{1-4}\cline{5-8}
$\Lambda_c^+\pi^+$ & 1.53  &      &      & $\Lambda_c^+\pi^+$  & Input  & $\Xi_c^0\pi^+$ & 1.54 \\
                   &       &      &      &                     &        & $\Xi_c^+\pi^0$ & 1.01 \\
\midrule[0.8pt]
\multicolumn{2}{c}{1.53}  & \multicolumn{2}{c}{$-$}  & \multicolumn{2}{c}{Input}  & \multicolumn{2}{c}{2.55} \\
\multicolumn{2}{c}{$1.89^{+0.09}_{-0.18}$~\cite{Agashe:2014kda}}  & \multicolumn{2}{c}{$-$}  & \multicolumn{2}{c}{$14.9\pm1.5$~\cite{Agashe:2014kda}}  & \multicolumn{2}{c}{$2.6\pm0.6$~\cite{Kato:2013ynr}} \\
\bottomrule[1pt]\bottomrule[1pt]
\end{tabular*}
\end{table}

\begin{table}[htbp]
\caption{Open-flavor strong decay widths of 1\emph{P} $\Lambda_c$ and $\Xi_c$ in MeV.}\label{table8}
\renewcommand\arraystretch{1.2}
\begin{tabular*}{85mm}{c@{\extracolsep{\fill}}ccccccc}
\toprule[1pt]\toprule[1pt]
                 \multicolumn{8}{c}{1\emph{P} $\Lambda_c$ and $\Xi_c$}   \\
\midrule[0.8pt]
\multicolumn{4}{c}{ $1/2^-$}  & \multicolumn{4}{c}{$3/2^-$} \\
\cline{1-4}\cline{5-8}
\multicolumn{2}{c}{$\Lambda_c(2595)^+$}  & \multicolumn{2}{c}{$\Xi_c(2790)^+$}  & \multicolumn{2}{c}{$\Lambda_c(2625)^+$}  & \multicolumn{2}{c}{$\Xi_c(2815)^+$} \\
\cline{1-4}\cline{5-8}
$\Sigma_c\pi$ & 2.78  & $\Xi'_c\pi$  & 6.01  & $\Sigma_c\pi$  & 0.04  & $\Xi'_c\pi$ & 0.15 \\
                   &       &      &      &                     &        & $\Xi^*_c\pi$ & 4.09 \\
\midrule[0.8pt]
\multicolumn{2}{c}{2.78}  & \multicolumn{2}{c}{6.01}  & \multicolumn{2}{c}{0.04}  & \multicolumn{2}{c}{4.24} \\
\multicolumn{2}{c}{$2.6\pm0.6$~\cite{Agashe:2014kda}}  & \multicolumn{2}{c}{$8.9\pm1.4$~\cite{Yelton:2016fqw}}  & \multicolumn{2}{c}{$<0.97$~\cite{Agashe:2014kda}}  & \multicolumn{2}{c}{$2.43\pm0.37$~\cite{Yelton:2016fqw}} \\
\bottomrule[1pt]\bottomrule[1pt]
\end{tabular*}
\end{table}

At present, all the ground states and 1\emph{P} $\mathcal{G}-$type charmed states have been experimentally established~\cite{Agashe:2014kda}. These states have been observed, at least, by two different collaborations, and their properties including masses and decays have been well determined. With good precision, the strong decays of these states provide a crucial test of our method.

Among the 1\emph{S} charmed baryons, the measurements of $\Sigma_c(2455)$ and $\Sigma_c(2520)$ have been largely improved~\cite{Aaltonen:2011sf,Lee:2014htd} (see Table \ref{table9}). In our calculation, the mass and decay width of $\Sigma_c(2520)^{++}$ measured by CDF will be taken as input data to fix the constant $\gamma$ peculiar to the $^3P_0$ model. With the transition factor for the process $\Sigma_c(2520)\rightarrow\Lambda_c(2286)+\pi$ (see Eq. (\ref{A12}) in the Appendix \ref{1}), the value of $\gamma$ is fixed as $1.296$.\footnote{For the different conventions to extract the color and flavor factors, the value of $\gamma$ here is different from those in Refs.~\cite{Chen:2007xf,Mu:2014iaa,Limphirat:2010zz}. The deviation, of course, dose not affect the predictions since $\gamma$ is regarded as an adjustable parameter in the $^3P_0$ model.}

\begin{table}[htbp]
\caption{The masses and widths (in units of MeV) of $\Sigma_c(2455)^{++}$ and $\Sigma_c(2520)^{++}$ measured by CDF~\cite{Aaltonen:2011sf} and Belle~\cite{Lee:2014htd}.
}\label{table9}
\renewcommand\arraystretch{1.2}
\begin{tabular*}{85mm}{c@{\extracolsep{\fill}}ccc}
\toprule[1pt]\toprule[1pt]
$\Sigma_c(2455)^{++}$ & 2453.90$\pm$0.13$\pm$0.14           & 2.34$\pm$0.47  & CDF  \\
                      & 2453.97$\pm$0.01$\pm$0.02$\pm$0.14  & 1.84$\pm$0.04$^{+0.07}_{-0.20}$  & Belle  \\
$\Sigma_c(2520)^{++}$ & 2517.19$\pm$0.46$\pm$0.14           & 15.03$\pm$2.52  & CDF   \\
                      & 2518.45$\pm$0.10$\pm$0.02$\pm$0.14  & 14.77$\pm$0.25$^{+0.18}_{-0.30}$  & Belle   \\
\bottomrule[1pt]\bottomrule[1pt]
\end{tabular*}
\end{table}

As shown in Tables \ref{table7} and \ref{table8}, the predicted widths of other 1\emph{S} charmed baryons are well consistent with experiments. Our results of mass spectra and decay widths indicate that $\Lambda_c(2595)^+$, $\Lambda_c(2625)^+$, $\Xi_c(2790)^{0,+}$, and $\Xi_c(2815)^{0,+}$ can be accommodated with the 1\emph{P} $\mathcal{G}-$type charmed baryons. $\Lambda_c(2595)^+$ and $\Xi_c(2790)^{0,+}$ can be classified into the $1/2^-$ states while $\Lambda_c(2625)^+$ and $\Xi_c(2815)^{0,+}$ into the $3/2^-$ states. The predicted mass splittings between the 1\emph{P} $1/2^-$ and $3/2^-$ states are 25 MeV and 27 MeV for the $\Lambda_c$ and $\Xi_c$ baryons, respectively, which are also consistent with the experiments. The assignments of $\Lambda_c(2595)^+$, $\Lambda_c(2625)^+$, $\Xi_c(2790)^{0,+}$, and $\Xi_c(2815)^{0,+}$ are also supported by other works~\cite{Ebert:2007nw,Ebert:2011kk,Chen:2009tm,Chen:2014nyo} in which the light quark cluster scenario was also employed. In addition, the mass spectra obtained by different types of the quark potential models in the three-body picture also support these assignments~\cite{Capstick:1986bm,Roberts:2007ni,Migura:2006ep,Garcilazo:2007eh,Shah:2016mig}. However, the investigations by QCD sum rules indicate that these 1\emph{P} candidates may have more complicated structures~\cite{Wang:2010hs,Wang:2015kua,Chen:2015kpa}. Especially, the work by Chen \emph{et al}. suggested that $\Lambda_c(2595)^+$ and $\Lambda_c(2625)^+$ form the heavy doublet $\tilde{\Lambda}_{c1}(1/2^-, 3/2^-)$ (the same assignments as the case of $\Xi_c(2790)^{0,+}$ and $\Xi_c(2815)^{0,+}$)~\cite{Chen:2015kpa}, which is different from our conclusion. Since the quantum numbers of $J^P$ have not yet been determined for these 1\emph{P} charmed states, more experiments are required in future.

\subsection{1\emph{P} $\Sigma_c^{~0,+,++}$ states}\label{b}

\begin{table}[htbp]
\caption{The partial and total decay widths of 1\emph{P} $\Sigma_c$ states in MeV.} \label{table10}
\renewcommand\arraystretch{1.2}
\begin{tabular*}{85mm}{@{\extracolsep{\fill}}lccccc }
\toprule[1pt]\toprule[1pt]
Decay  &\multicolumn{2}{c}{$1/2^-~(1P)$}  & \multicolumn{2}{c}{$3/2^-~(1P)$}  & $5/2^-~(1P)$  \\
\cline{2-3}\cline{4-5}\cline{6-6}
modes  & $\Sigma_{c0}(2702)$  & $\Sigma_{c1}(2765)$  & $\Sigma_{c1}(2798)$ &$\Sigma_{c2}(2785)$ & $\Sigma_{c2}(2790)$   \\
\midrule[0.8pt]
 $\Lambda_c\pi$       &  3.64    & $\times$ & $\times$  & 24.06  &  24.63    \\
 $\Sigma_c(2455)\pi$  & $\times$ & 58.94    & 3.48      & 5.22   &  2.50     \\
 $\Sigma_c(2520)\pi$  & $\times$ & 1.70     & 63.72     & 2.47   &  4.34       \\
 $\Lambda_c(2595)\pi$ &          & 2.88     & 2.31      & 1.93   &  0.03       \\
 $\Lambda_c(2625)\pi$ &          &          & 3.12      & 0.07   &  0.63      \\
 \midrule[0.8pt]
  Theory              & 3.64     &  63.52   & 72.63     & 33.75  &  32.13      \\
  Expt.~\cite{Agashe:2014kda} &          & $\approx50$ &        & $72^{+22}_{-15}$    &       \\
\bottomrule[1pt]\bottomrule[1pt]
\end{tabular*}
\end{table}

As shown in Tables \ref{table2} and \ref{table3}, the masses of 1\emph{P} $\Sigma_c$ states are predicted in the range of 2700$\sim$2800 MeV. Then, $\Sigma_c(2765)^+$ and $\Sigma_c(2800)^{0,+,++}$ can be grouped into the candidates of 1\emph{P} $\Sigma_c$ family. The predicted mass of $|1P,1/2^-\rangle_h$ state is about 2765 MeV which is in good agreement with the measured mass of $\Sigma_c(2765)^+$. In addition, the theoretical result for the decay width of the $\Sigma_{c1}(1/2^-)$ state in Table \ref{table10} is about 63.52 MeV which is also in agreement with the measurements~\cite{Artuso:2000xy, Abe:2006rz,Joo:2014fka}. Furthermore, the signal of $\Sigma_c(2765)^+$ has been observed in the $\Sigma_c(2455)\pi$ intermediate state while there is no clear evidence for the decay of $\Sigma_c(2765)^+$ through $\Sigma_c(2520)\pi$~\cite{Abe:2006rz,Joo:2014fka}. This is also consistent with our results of the $|1P,1/2^-\rangle_h$. Based on the combined analysis of the mass spectrum and strong decays, we, therefore, conclude that $\Sigma_c(2765)^+$ could be regarded as a good candidate of $\Sigma_{c1}(1/2^-)$. Considering uncertainties of the quark potential models, the masses obtained by Refs~\cite{Ebert:2007nw,Ebert:2011kk,Capstick:1986bm} are not contradictory to our assignment to $\Sigma_c(2765)^+$.

According to the predicted masses in Table \ref{table3}, $\Sigma_c(2800)^{0,+,++}$ could be assigned to either $|1P,3/2^-\rangle_l$, or $|1P,3/2^-\rangle_h$, or $|1P,5/2^-\rangle$ states. When we consider the decay properties of these three states (see Table \ref{table10}), the possibility of assignment to the $|1P,3/2^-\rangle_l$ state can be excluded since the Belle Collaboration observed this state in the $\Lambda_c^+\pi$ mode.\footnote{Even the possible mixing between $\Sigma_{c1}(3/2^-)$ and $\Sigma_{c2}(3/2^-)$ is considered, the partial width of $\Lambda_c^+\pi$ is only 3.87 MeV for the $|1P,3/2^-\rangle_l$ state where the mixing angle obtained in Table \ref{table4} has been used.} At present, the Belle Collaboration tentatively identified $\Sigma_c(2800)^{0,+,++}$ as members of the $\Sigma_{c2}(3/2^-)$ isospin triplet, which agrees with our results of both mass spectrum and strong decays. When the measured mass of $\Sigma_c(2800)^0$ (2806 MeV) is used for the $\Sigma_{c2}(3/2^-)$ state, the predicted width is about 40.1 MeV which is comparable with the experiment~\cite{Mizuk:2004yu}. However, we notice that the quantum number $J^P$ of $\Sigma_c(2800)^{0,+,++}$ has not yet been measured. Then the possibility of this state as the $\Sigma_{c2}(5/2^-)$ candidate can not be excluded by our results since the decay mode of $\Lambda_c^+\pi$ is dominant for this state. In addition, the predicted mass and total width of $\Sigma_{c2}(5/2^-)$ state are also compatible with experimental data of the $\Sigma_c(2800)^{0,+,++}$ baryon. Therefore, we would like to point out that the signal of $\Sigma_c(2800)^{0,+,++}$ found by Belle might be their overlapping structure. We hope the future experiments measure the following branching ratios to disentangle this state:

For~the~$\Sigma_{c2}(3/2^-)$~state,
\begin{equation}
\frac{\mathcal {B}(\Sigma_{c2}(3/2^-)\rightarrow \Sigma_c(2455)~\pi)}{\mathcal
{B}(\Sigma_{c2}(3/2^-)\rightarrow \Sigma_c(2520)~\pi)}~=~1.90; \label{eq22}
\end{equation}

\begin{equation}
\frac{\mathcal {B}(\Sigma_{c2}(3/2^-)\rightarrow \Lambda_c(2286)~\pi)}{\mathcal
{B}(\Sigma_{c2}(3/2^-)\rightarrow \Sigma_c(2455)~\pi)}~=~4.07 \label{eq23}
\end{equation}

For~the~$\Sigma_{c2}(5/2^-)$~state,
\begin{equation}
\frac{\mathcal {B}(\Sigma_{c2}(5/2^-)\rightarrow \Sigma_c(2455)~\pi)}{\mathcal
{B}(\Sigma_{c2}(5/2^-)\rightarrow \Sigma_c(2520)~\pi)}~=~0.58. \label{eq24}
\end{equation}

\begin{equation}
\frac{\mathcal {B}(\Sigma_{c2}(5/2^-)\rightarrow \Lambda_c(2286)~\pi)}{\mathcal
{B}(\Sigma_{c2}(5/2^-)\rightarrow \Sigma_c(2455)~\pi)}~=~9.85  \label{eq25}
\end{equation}

As mentioned earlier, the signal $\Sigma_c(2850)^0$ discovered by the {BaBar} collaboration may be a $J=1/2$ state. If $\Sigma_c(2850)^0$ is the $1/2^+(2S)$ state, the corresponding ratios (see Subsection~\ref{e}) are different from Eqs.~(\ref{eq22}$\sim$\ref{eq25}). So the measurements of these ratios of branching fractions can help us understand the nature of $\Sigma_c(2800)^{0,+,++}$ and $\Sigma_c(2850)^0$.

Although, at present, the $\Sigma_{c0}(1/2^-)$ and $\Sigma_{c1}(3/2^-)$ states are still missing in experiments, our results indicate that the $\Sigma_{c0}(1/2^-)$ state may be a narrow resonance and its predominant decay channel $\Lambda_c^+\pi$ is only about 3.64 MeV (see Table \ref{table10}). Since the decay mode of $\Sigma_c(2520)~\pi$ is the largest for the $\Sigma_{c1}(3/2^-)$ state, we suggest to search this channel for this state in the future experiments. In the heavy quark limit, the following branching ratio 
for $\Sigma_{c1}(3/2^-)$~state
\begin{eqnarray}
\frac{\mathcal {B}(\Sigma_{c1}(3/2^-)\rightarrow \Sigma_c(2455)~\pi)}{\mathcal
{B}(\Sigma_{c1}(3/2^-)\rightarrow \Sigma_c(2520)~\pi)}~=~0.05. \label{eq26}
\end{eqnarray}
is much smaller than $\Sigma_{c2}(3/2^-)$ (Eq.~\ref{eq22}).

\subsection{1\emph{P} $\Xi_c^{\prime~0,+}$ states}\label{c}

\begin{table}[htbp]
\caption{The partial and total decay widths of 1\emph{P} $\Xi^\prime_c$ states in MeV.} \label{table11}
\renewcommand\arraystretch{1.2}
\begin{tabular*}{86mm}{@{\extracolsep{\fill}}lccccc}
\toprule[1pt]\toprule[1pt]
Decay  &\multicolumn{2}{c}{$1/2^-~(1P)$}  & \multicolumn{2}{c}{$3/2^-~(1P)$}  & $5/2^-~(1P)$  \\
\cline{2-3}\cline{4-5}\cline{6-6}
modes  & $\Xi'_{c0}(2839)$  & $\Xi'_{c1}(2900)$  & $\Xi'_{c1}(2932)$ & $\Xi'_{c2}(2921)$ & $\Xi'_{c2}(2927)$   \\
\midrule[0.8pt]
 $\Lambda_cK$      &  46.59   & $\times$ & $\times$  & 11.59  &  12.43    \\
 $\Xi_c\pi$        &  4.39    & $\times$ & $\times$  & 7.42   &  7.75     \\
 $\Xi'_c(2580)\pi$ & $\times$ & 9.44     & 0.76      & 1.20   &  0.57       \\
 $\Xi'_c(2645)\pi$ & $\times$ & 0.52     & 3.23      & 0.75   &  1.31       \\
 $\Xi_c(2790)\pi$  &          &          & 0.01      &        &         \\
 \midrule[0.8pt]
  Theory           & 50.98    &  9.96    &  4.00     & 20.96   &  22.06      \\
  Expt.            &          &          &           & \multicolumn{2}{c}{$36\pm7\pm11$~\cite{Aubert:2007eb}}       \\
\bottomrule[1pt]\bottomrule[1pt]
\end{tabular*}
\end{table}

As shown in Table \ref{table3}, the predicted masses of 1\emph{P} $\Xi_c^\prime$ is in the range from 2840 to 2930 MeV. Then the resonance structure observed by BaBar~\cite{Aubert:2007eb} in the decay channel $B^-\rightarrow\Xi_c^{\prime}(2930)^0\bar{\Lambda}_c^-\rightarrow\Lambda^+_cK^-\bar{\Lambda}_c^-$ with an invariant mass of 2.93 GeV could be a good candidate of 1\emph{P} $\Xi_c^\prime$ members. The results of decays in Table \ref{table11} favor $\Xi_c^{\prime}(2930)^0$ as the $\Xi'_{c2}(3/2^-)$ or $\Xi'_{c2}(5/2^-)$ state. Then $\Xi_c^{\prime}(2930)^0$ might be regarded as the strange partner of $\Sigma_c(2800)^{0,+,++}$ by our results. Interestingly, the mass difference between $\Xi_c^{\prime}(2930)^0$ and $\Sigma_c(2800)^{0,+,++}$ is about 130 MeV which is comparable with the mass differences among sextet states of ground charmed baryons~\cite{Cheng:2015naa}. With a chiral quark model, Liu \emph{et al}. also analyzed the $\Xi_c^{\prime}(2930)^0$ by the two-body strong decays~\cite{Liu:2012sj}. Their results support $\Xi_c^{\prime}(2930)^0$ as the $|\Xi_c^{\prime2}P_\lambda,~1/2^-\rangle$ or $|\Xi_c^{\prime4}P_\lambda,~1/2^-\rangle$ state. Since the heavy quark symmetry was not considered in Ref.~\cite{Liu:2012sj}, the notations of charm-strange baryons in Ref~\cite{Liu:2012sj} are different from our $\Xi'_{c0}(1/2^-)$ and $\Xi'_{c1}(1/2^-)$. Although the results in Table~\ref{table11} indicate that the $\Lambda_c^+K$ decay mode dominates the decay of $\Xi^\prime_{c0}(1/2^-)$ state, the mass of this state is predicted about 2840 MeV which is much smaller than $\Xi_c^{\prime0}(2930)$. In addition, the $\Lambda_c^+K$ decay mode is forbidden for the $\Xi^\prime_{c1}(1/2^-)$ state. Thus, according to our results, $\Xi_c^{\prime0}(2930)$ is unlikely to be a 1\emph{P} state with $J^P=1/2^-$.

Another charm-strange baryon, $\Xi_c(2980)^{0,+}$, is slightly higher than the predicted mass range of 1\emph{P} $\Xi_c^\prime$ states. This state has been observed in $\Sigma_c(2455)K$, $\Xi^\prime_c(2580)\pi$, $\Xi^\prime_c(2645)\pi$, and nonresonant $\Lambda^+_c\bar{K}\pi$ decay channels. However, it was not seen in the decay modes of $\Lambda^+_c\bar{K}$ and $\Xi_c\pi$~\cite{Chistov:2006zj,Aubert:2007dt,Lesiak:2008wz}. Comparing the mass and decay properties of $\Xi_c(2980)^{0,+}$ with our results, the possibility as a 1\emph{P} $\Xi^\prime_c$ state might be excluded. As shown in the next Subsection, $\Xi_c(2980)^{0,+}$ could be a good 2\emph{S} $\Xi_c$ candidate. Based on our results on strong decays, we find that the $\Xi^\prime_{c1}(1/2^-)$ and $\Xi^\prime_{c1}(3/2^-)$ are quite narrow (see Table \ref{table11}).

\subsection{2\emph{S} $\Lambda_c^+$ and $\Xi_c^{0,+}$ states}\label{d}

According to the mass spectrum (see Table~\ref{table2}), $\Lambda_c/\Sigma_c(2765)^+$ can also be regarded as the first radial (2\emph{S}) excitation of the $\Lambda_c(2286)^+$ with $J^P=1/2^+$. Interestingly, the results of strong decays in Table \ref{table12} do not contradict with this assignment. Our calculation indicates that the decay channel $\Sigma_c(2455)\pi$ is a dominant decay channel for the $\Lambda_c^+(2S)$ state. This is in line with the observations by Belle~\cite{Abe:2006rz,Joo:2014fka}. At present, both $1/2^+(2S)$ $\Lambda_c^+$ and $1/2^-(1P)$ $\Sigma_c^+$ are possible for the assignment of $\Lambda_c/\Sigma_c(2765)^+$. However, there is a very important feature for experiments to distinguish these two assignments in future. Specifically, we suggest to search $\Lambda_c/\Sigma_c(2765)^+$ in the channel of $\Sigma_c(2520)\pi$. As shown in Table~\ref{table12}, the channel $\Sigma_c(2520)\pi$ is large enough to find the $\Lambda_c^+(2S)$ state. On the other hand, this mode seems too small to be detected for the $\Sigma_{c1}(1/2^-)$(see Table~\ref{table10}). Explaining the criteria concretely, we give the following branching ratios for these two states,

For~the~$\Lambda_c(2S)$~state,
\begin{eqnarray}
\frac{\mathcal {B}(\Lambda_c(2765)\rightarrow \Sigma_c(2520)~\pi)}{\mathcal
{B}(\Lambda_c(2765)\rightarrow \Sigma_c(2455)~\pi)}~=~0.74.  \label{eq27}
\end{eqnarray}

For~the~$\Sigma_{c1}(1/2^-)$~state,
\begin{eqnarray}
\frac{\mathcal {B}(\Sigma_c(2765)\rightarrow \Sigma_c(2520)~\pi)}{\mathcal
{B}(\Sigma_c(2765)\rightarrow \Sigma_c(2455)~\pi)}~=~0.03.   \label{eq28}
\end{eqnarray}
The branching ratio of $\mathcal {B}(\Sigma_c(2520)\pi)/\mathcal {B}(\Sigma_c(2455)\pi)$ for the $\Sigma_{c1}(1/2^-)$ state is roughly an order of magnitude smaller than $\Lambda_c(2S)$. If $\Lambda_c(2765)$ is the 2\emph{S} excitation, $\Xi_c(2980)$ could be a good candidate as its charm-strange analog~\cite{Cheng:2015naa} as seen in Table~\ref{table12}. The mass difference between $\Lambda_c(2765)$ and $\Xi_c(2980)$ is about 200 MeV which nearly equals the mass difference between $\Lambda_c(2287)$ and $\Xi_c(2470)$. The predicted width of $\Xi_c(2980)$ is 27.44 MeV which is in good agreement with the experimental data~\cite{Yelton:2016fqw,Aubert:2007dt}. As the 2\emph{S} excitation of $\Xi_c(2470)$, the branching ratio,
\begin{eqnarray}
\frac{\mathcal {B}(\Xi_c(2980)\rightarrow \Xi^\prime_c(2580)~\pi)}{\mathcal
{B}(\Xi_c(2980)\rightarrow \Xi_c(2645)~\pi)}~=~0.89,   \label{eq29}
\end{eqnarray}
is predicted for $\Xi_c(2980)$, which can be tested by the future experiments. Recently, the following ratio of branching fractions
\begin{eqnarray}
\frac{\mathcal {B}(\Xi_c(2980)^+\rightarrow\Xi^\prime_c(2580)^0\pi^+)}{\mathcal
{B}(\Xi_c(2815)^+\rightarrow \Xi_c(2645)^0\pi^+\rightarrow\Xi_c^+\pi^-\pi^+)}~\approx~75\%,  \label{eq30}
\end{eqnarray}
has been estimated by the Belle Collaboration~\cite{Yelton:2016fqw}. Combining this with the predicted partial widths of $\Xi_c(2815)$ and $\Xi_c(2645)$ in Tables~\ref{table7} and \ref{table8}, the branching fraction $\mathcal {B}(\Xi_c(2980)^+\rightarrow\Xi^\prime_c(2580)^0\pi^+)$ is evaluated about $40\%$ which is consistent with our direct result of $41.8\%$.

\begin{table*}[htbp]
\caption{The partial and total decay widths of 2\emph{S} $\Lambda_c^+$ and $\Xi_c^{+,0}$ states in MeV.} \label{table12}
\renewcommand\arraystretch{1.2}
\begin{tabular*}{175mm}{@{\extracolsep{\fill}}lccccccccccc}
\toprule[1pt]\toprule[1pt]
\multicolumn{4}{c}{$1/2^+~(2S)$}  & \multicolumn{4}{c}{$1/2^+~(2S)'$}  & \multicolumn{4}{c}{$3/2^+~(2S)'$}  \\
\cline{1-4}\cline{5-12}
\multicolumn{2}{c}{$\Lambda_c(2765)^+$}& \multicolumn{2}{c}{$\Xi_c(2980)$}  & \multicolumn{2}{c}{$\Sigma_c(2850)^0$}&\multicolumn{2}{c}{$\Xi'_c(3000)$}& \multicolumn{2}{c}{$\Sigma_c(2880)^0$}& \multicolumn{2}{c}{$\Xi'_c(3030)$}  \\
\cline{1-4}\cline{5-8}\cline{9-12}
$\Sigma_c(2455)\pi$ &  26.23  & $\Sigma_c(2455)K$ & 3.14  & $\Lambda_c^+\pi$       &  35.11  & $\Lambda_c^+K$   & 17.42 & $\Lambda_c^+\pi$     & 34.96 & $\Lambda_c^+K$       &  18.37   \\
$\Sigma_c(2520)\pi$ &  19.28  & $\Xi'_c(2580)\pi$ & 11.47 & $\Sigma_c(2455)\pi$  &  57.16  & $\Xi'_c(2580)\pi$  & 12.56 & $\Sigma_c(2455)\pi$  & 15.98 & $\Xi'_c(2580)\pi$  &  3.50   \\
                    &         & $\Xi'_c(2645)\pi$ & 12.83 & $\Sigma_c(2520)\pi$  &  17.54  & $\Xi'_c(2645)\pi$  & 4.13  & $\Sigma_c(2520)\pi$  & 54.52 & $\Xi'_c(2645)\pi$  &  12.92  \\
                    &         &                   &       & $\Lambda_c(2595)\pi$ &  6.92   & $\Xi_c(2790)\pi$   & 5.89  & $\Lambda_c(2595)\pi$ & 1.07  & $\Xi_c(2790)\pi$   &  0.40  \\
                    &         &                   &       & $\Lambda_c(2625)\pi$ &  1.57   & $\Xi_c(2815)\pi$   & 0.13  & $\Lambda_c(2625)\pi$ & 7.62  & $\Xi_c(2815)\pi$   &  6.22   \\
                    &         &                   &       & $D^0n$               &  0.03   & $\Sigma_c(2455)K$  & 15.34 & $D^0n$               &  3.03 & $\Sigma_c(2455)K$  &  6.49   \\
                    &         &                   &       &                      &         &                    &       &                      &       & $D^0\Lambda$               &  0.01  \\
\midrule[0.8pt]
\multicolumn{2}{c}{45.51}& \multicolumn{2}{c}{27.44}  & \multicolumn{2}{c}{118.33}&\multicolumn{2}{c}{55.47}& \multicolumn{2}{c}{117.18}& \multicolumn{2}{c}{47.91}  \\
\multicolumn{2}{c}{$\approx50$~\cite{Agashe:2014kda}}& \multicolumn{2}{c}{28.1$\pm2.4^{+1.0}_{-5.0}$~\cite{Yelton:2016fqw}}  & \multicolumn{2}{c}{86$^{+33}_{-22}$~\cite{Aubert:2008ax}}&\multicolumn{2}{c}{ }& \multicolumn{2}{c}{ }& \multicolumn{2}{c}{ }  \\
\bottomrule[1pt]\bottomrule[1pt]
\end{tabular*}
\end{table*}

\subsection{2\emph{S} $\Sigma_c^{0,+,++}$ and $\Xi_c^{\prime~0,+}$ states}\label{e}

In Table~\ref{table3}, masses of the 2\emph{S} $\Sigma_c(1/2^+,3/2^+)$ states are predicted as 2850 MeV and 2876 MeV, respectively. The neutral $\Sigma_c(2850)^0$ found by the BaBar Collaboration in the decay channel $B^-\rightarrow\Sigma_c(2850)^0\bar{p}\rightarrow\Lambda_c^+\pi^-\bar{p}$~\cite{Aubert:2008ax} can be regarded as the 2\emph{S} $\Sigma_c$ state with $J^P=1/2^+$. The mass and width of the neutral $\Sigma_c(2800)^0$ and $\Sigma_c(2850)^0$ are collected below.
\begin{equation}
\begin{split}
&\Sigma_c(2800)^0:~~~m~=~2806^{+5}_{-7}~\rm {MeV},~~~\Gamma~=~72^{+22}_{-15}~\rm {MeV};\\ &
\Sigma_c(2850)^0:~~~m~=~2846\pm8\pm10~\rm {MeV},~~~\Gamma~=~86^{+33}_{-22}~\rm {MeV}.\nonumber
\end{split}
\end{equation}

For lack of experimental information, at present, PDG treated $\Sigma_c(2850)^0$ and $\Sigma_c(2800)^{0,+,++}$ as the same state~\cite{Agashe:2014kda}. As pointed out by the BaBar collaboration~\cite{Aubert:2008ax}, however, there are indications that these two signals detected by Belle~\cite{Mizuk:2004yu} and BaBar~\cite{Aubert:2008ax} are two different $\Sigma_c^*$ states. The main reasons are listed as follows:
\begin{enumerate}
\item Although the widths of $\Sigma_c(2800)^{0,+,++}$ and $\Sigma_c(2850)^0$ are consistent with each other, their masses are $3\sigma$ apart.

\item The Belle Collaboration tentatively identified the $\Sigma_c(2800)^{0,+,++}$ as the $J=3/2$ isospin triple, while the BaBar Collaboration found the weak evidence of $\Sigma_c(2850)^0$ as a $J=1/2$ state.

\end{enumerate}

Our results also indicate that $\Sigma_c(2800)^{0,+,++}$ and $\Sigma_c(2850)^0$ are the different $\Sigma_c$ excited states. One notices that the predicted mass of $1/2^+(2S)$ $\Sigma_c$ state in this work and in Ref.~\cite{Ebert:2007nw} are around 2850 MeV. Even the results in Refs.~\cite{Ebert:2011kk,Capstick:1986bm} are only about 50 MeV larger than the measurements. Due to the intrinsic uncertainties of the quark potential model, it is appropriate to assign $\Sigma_c(2850)^0$ as a 2\emph{S} $1/2^+$ state. More importantly, the predicted decay width of $\Sigma_c(1/2^+,~2S)$ state is 118.36 MeV which is comparable with the measurement by BaBar~\cite{Aubert:2008ax}. The partial width of $\Lambda_c\pi$ is 35.11 MeV, which can explain why $\Sigma_c(2850)^0$ was first found in this channel. We find that the decay modes of $\Sigma_c(2455)\pi$ and $\Sigma_c(2520)\pi$ are also large. Finally, we give the following branching ratios,
\begin{eqnarray}
\frac{\mathcal {B}(\Sigma_c(2850)\rightarrow\Sigma_c(2455)~\pi)}{\mathcal
{B}(\Sigma_c(2850)\rightarrow\Sigma_c(2520)~\pi)}~=~3.26   \label{eq31}
\end{eqnarray}
and
\begin{eqnarray}
\frac{\mathcal {B}(\Sigma_c(2850)\rightarrow\Sigma_c(2286)~\pi)}{\mathcal
{B}(\Sigma_c(2850)\rightarrow\Sigma_c(2455)~\pi)}~=~0.61,   \label{eq32}
\end{eqnarray}
which can be tested by future experiments. If $\Sigma_c(2850)^0$ is the $1/2^+(2S)$ state, the mass of its doublet partner in the heavy quark effective theory is predicted as 2876 MeV (denoted as $\Sigma_c(2880)$). According to the predicted decay widths in Table \ref{table12}, this state might also be broad. $\Lambda_c^+\pi$, $\Sigma_c(2455)\pi$, and $\Sigma_c(2520)\pi$ are also dominant for the decay of $\Sigma_c(2880)$. The ratio of $\Gamma(\Sigma_c(2455)\pi)/\Gamma(\Sigma_c(2520)\pi)$ for $\Sigma_c(2880)$ is different from $\Sigma_c(2850)$, whose numerical value is given by,
\begin{eqnarray}
\frac{\mathcal {B}(\Sigma_c(2880)\rightarrow\Sigma_c(2455)~\pi)}{\mathcal
{B}(\Sigma_c(2880)\rightarrow\Sigma_c(2520)~\pi)}~=~0.29.   \label{eq33}
\end{eqnarray}
Even though the strange partners of $\Sigma_c(2850)$ and $\Sigma_c(2880)$ have not been found by any experiments, their decay properties are calculated and presented in Table \ref{table12}. Our results indicate that $\Lambda_c^+K$, $\Xi_c^\prime(2580)\pi$, and $\Sigma_c(2455)K$ are the dominant decay modes of the $\Xi_c^\prime(3000)$ state with $J^P=1/2^+$, while $\Lambda_c^+K$ and $\Xi_c^\prime(2645)\pi$ are those of the $\Xi_c^\prime(3030)$.\footnote{If $\Xi_c(2980)$ is the first radial excited state of $\Xi_c(2470)$. Then our predicted masses for 2\emph{S} charm-strange baryons may be about 20$\sim$30 MeV lower than experiments. To compensate this difference, we increase about 25 MeV for the 2\emph{S} $\Xi_c^\prime$ statesm in this case.} Besides the masses and decay widths, the following branching ratios may also be valuable for future experiments:
\begin{eqnarray}
\frac{\mathcal {B}(\Xi^\prime_c(3000)\rightarrow\Xi^\prime_c(2580)~\pi)}{\mathcal
{B}(\Xi^\prime_c(3000)\rightarrow\Xi^\prime_c(2645)~\pi)}~=~3.04,   \label{eq34}
\end{eqnarray}
and
\begin{eqnarray}
\frac{\mathcal {B}(\Xi^\prime_c(3030)\rightarrow\Xi^\prime_c(2580)~\pi)}{\mathcal
{B}(\Xi^\prime_c(3030)\rightarrow\Xi^\prime_c(2645)~\pi)}~=~0.27.   \label{eq35}
\end{eqnarray}

\section{Summary and Outlook}\label{sec5}
In principle, both $\rho$ and $\lambda$ modes can be excited in a baryon system. For charmed baryons, the excitation energies of the $\rho$ and $\lambda$ modes are different due to the heavier mass of a \emph{c} quark. For the ordinary confining potential, such as the linear or harmonic form, the excited energy of the $\rho$ mode is larger than the $\lambda$ mode~\cite{Copley:1979wj}. Hence the low excited charmed baryons may be 
dominated by the $\lambda$ mode excitations. Recently, the investigation by Yoshida \emph{et~al}. confirmed this point~\cite{Yoshida:2015tia}. Furthermore, they find that the $\rho$ and $\lambda$ modes are well separated for the charmed and bottom baryons, which means the component of the $\rho$ mode can be ignored for the low excited charmed baryons. Interestingly, the works~\cite{Ebert:2007nw,Ebert:2011kk,Chen:2014nyo} have also shown that the masses of existing charmed baryons can be explained by the $\lambda$ mode. Hence, our study of strong decays of the low excited charmed baryons is an important complement to these works~\cite{Ebert:2007nw,Ebert:2011kk,Chen:2014nyo,Yoshida:2015tia}.

Up to now, several candidates of the 1\emph{P} and 2\emph{S} charm and charm-strange baryons have been found by experiments, and some of them are still open to debate. To better understand these low excited charmed baryons, in this paper, we carry a systematical study of the mass spectra and strong decays for the 1\emph{P} and 2\emph{S} charmed baryon states in the framework of the nonrelativistic constituent quark model. The masses have been calculated in the potential model where the charmed baryons are simply treated as a quasi two body system in a light quark cluster picture. The strong decays are computed by the EHQ decay formula where the transition factors are determined by the $^3P_0$ model. When calculating the decays, the inner structure of a light quark cluster has also been considered. Except for the unique parameter $\gamma$ of the QPC model, the parameters in the potential model and in the EHQ decay formula have the same values.

The well-established ground and 1\emph{P} $\mathcal{G}-$type charmed baryons provide a good test to our method. The experimental properties including both masses and widths for these states can be well explained by our results. This success has made us more confident of our predictions for other 1\emph{P} and 2\emph{S} states. Our main conclusions are given as follows:

The broad state $\Lambda_c(2765)^+$ (or $\Sigma_c(2765)^+$) which is still ambiguous could be assigned to the $1/2^+(2S)$ $\Lambda_c^+$, or the $1/2^-(1P)$ $\Sigma_{c1}^+$ state. The branching ratio $\mathcal{B}(\Sigma_c(2455)\pi)/\mathcal{B}(\Sigma_c(2520)\pi)$ is found to be different for these two assignments, which may help us understand the nature of this state.

$\Sigma_c(2800)^{0,+,++}$ observed by the Belle Collaboration in $e^+e^-$ annihilation processes~\cite{Mizuk:2004yu} can be regarded as a negative parity state with $J^P=3/2^-$, or $5/2^-$, or their overlapping structure. We suggest to measure the $\mathcal{B}(\Sigma_c(2455)\pi)/\mathcal{B}(\Sigma_c(2520)\pi)$ in future. Another neutral state, $\Sigma_c(2850)^0$, which was found in the $B^-$ meson decay~\cite{Aubert:2008ax} could be a good candidate for the first radial excited state of $\Sigma_c(2455)$. With the above assignments, the ratios of $\mathcal{B}(\Lambda_c(2287)\pi)/\mathcal{B}(\Sigma_c(2455)\pi)$ shall be very different for $\Sigma_c(2800)^{0,+,++}$ and $\Sigma_c(2850)^0$, \emph{i.e.}, 4.07 for $\Sigma_c(2800)^{0,+,++}$ and 0.61 for $\Sigma_c(2850)^0$. The puzzle of $\Sigma_c(2800)^{0,+,++}$ and $\Sigma_c(2850)^0$ may be disentangled if these branching ratios are measured in future. In addition, the ratio of branching fractions $\mathcal{B}(\Sigma_c(2455)\pi)/\mathcal{B}(\Sigma_c(2520)\pi)$ for $\Sigma_c(2850)^0$ is predicted to be 3.26.

The analysis of the mass and decay properties supports that $\Xi_c(2980)^{0,+}$ is the 2\emph{S} excitation (the first radial excited state of $\Xi_c(2470)$). The existence of $\Xi_c(2930)^0$ is still in dispute. If it exists, the assignments of $\Xi_{c2}^\prime(3/2^-)$ and $\Xi_{c2}^\prime(5/2^-)$ are possible. In other words, it could be regarded as a strange partner of $\Sigma_c(2800)^{0,+,++}$. Some useful ratios of partial decay widths are also presented for $\Xi_c(2980)^{0,+}$ and $\Xi_c(2930)^0$.

Although both the masses and strong decays have been explained in the heavy quark-light quark cluster picture for the observed 2\emph{S} and 1\emph{P} candidates, it is not the end of the story to study the excited charmed baryon states. Investigation of the $\rho$ mode excited states with higher energies are also important to identify the effective degrees of freedom of charmed baryons. However, this topic needs much laborious work and is beyond the scope of the present work. In addition, the quark model employed here neglects the effect of virtual hadronic loops. In future, a more reasonable scheme for studying the properties of heavy baryons will be obtained by the unquenched quark model. Another topic which is left as a future task is to calculate the sum rules among the branching fractions of charmed baryons by applying the technique found in Ref.~\cite{Matsuki:2011xp}.

\section*{Acknowledgement}

Bing Chen thanks to Franz F. Sch\"oberl for the package which is very useful to solve the Schr$\rm {\ddot{o}}$dinger equation. This project is supported by the National Natural Science Foundation of China under Grant Nos. 11305003, 11222547, 11175073, 11447604 and U1204115. Xiang Liu is also supported by the National  Program for Support of Top-notch Young Professionals.

\appendix
\section{Transition factor $\mathcal {M}^{j_A,j_B}_{j_C,\ell}(q)$ in the QPC model}\label{1}

In the following, we will show how to obtain the partial wave amplitudes by the $^3P_0$ strong decay model for the decays of excited charmed baryons. As an example, the process $\Sigma_c(2520)\rightarrow\Lambda_c(2280)\pi$ will be constructed and the transition factor for the EHQ formula will be extracted.

As pointed in Section \ref{sec3}, there are two possible decay processes for an excited charmed baryon state (see Fig.~\ref{Fig3}). The final states of the left figure contain a charmed baryon and a light meson. The right one contains a charmed meson and a light baryon. If a baryon decays via the so-called $^3P_0$ mechanism, a quark-antiquark pair is created from the vacuum and then regroups two outgoing hadrons by a quark rearrangement process. In the non-relativistic limit, the transition operator $\mathcal {\hat{T}}$ of the $^3P_0$ model is given by
\begin{equation}
\begin{split}
\mathcal {\hat{T}}=&-3\gamma
\sum_{\text{\emph{m}}}\langle1,m;1,-m|0,0\rangle \iint
d^3\vec{k}_4d^3\vec{k}_5\delta^3(\vec{k}_4+\vec{k}_5)\\ &\times\mathcal
{Y}_1^m(\frac{\vec{k}_4-\vec{k}_5}{2})\omega^{(4,5)}\varphi^{(4,5)}_0\chi^{(4,5)}_{1,-m}d^\dag_{4}(\vec{k}_4)d^\dag_{5}(\vec{k}_5),
\end{split}\label{A1}
\end{equation}
where the $\omega_0^{(4,5)}$ and $\varphi^{(4,5)}_0$ are the color and flavor wave functions of the $q_4\bar{q}_5$ pair created from the vacuum. Thus,
$\omega^{(4,5)}=(R\bar{R}+G\bar{G}+B\bar{B})/\sqrt{3}$ and $\varphi^{(4,5)}_0=(u\bar{u}+d\bar{d}+s\bar{s})/\sqrt{3}$ are color and flavor singlets. The pair is also assumed to carry the quantum number of $0^{++}$, suggesting that they are in a $^3P_0$ state. The $\chi^{(4,5)}_{1,-m}$ represents the pair production in a spin
triplet state. The solid harmonic polynomial $\mathcal {Y}_1^m(\vec{k})\equiv|\vec{k}|\mathcal {Y}_1^m(\theta_k,\phi_k)$ reflects the momentum-space distribution of the $q_4\bar{q}_5$. $\gamma$ is a dimensionless constant which expresses the strength of the quark-antiquark pair created from the vacuum. The value of $\gamma$ is usually fixed by fitting the well measured partial decay widths.

When the mock state~\cite{Hayne:1981zy} is adopted to describe the spatial wave function of a meson, the helicity amplitude $\mathcal {M}^{j_A,j_B,j_C}(q)$ can be easily constructed in the $LS$ basis~\cite{LeYaouanc:1988fx}. The mock state for an \emph{A} meson is

\begin{equation}
\begin{split}
|A({n_A}^{2S_A+1}&L_A^{J_Aj_A}(\vec{P}_A)\rangle\equiv\\
&\omega_A^{123}\phi_A^{123}\prod_A\int d^3\vec{k}_1d^3\vec{k}_2d^3\vec{k}_3\delta^3(\vec{k}_1+\vec{k}_2+\vec{k}_3-\vec{P}_A)\\
&\times\Psi_{n_A}^{L_Al_A}(\vec{k}_1,\vec{k}_2,\vec{k}_3)|q_1(\vec{k}_1)q_2(\vec{k}_2)q_3(\vec{k}_3)\rangle.
\end{split}\label{A2}
\end{equation}
As for the left decay process in Fig.~\ref{Fig3}, the wave function of a \emph{B} baryon can be constructed in the same way. The wave function of a \emph{C} meson is
\begin{equation}
\begin{split}
|C({n_C}^{2S_C+1}&L_C^{J_Cj_C}(\vec{P}_C)\rangle\equiv\omega_C^{15}\phi_C^{15}\prod_C \int d^3\vec{k}_1d^3\vec{k}_5\\
&\times\delta^3(\vec{k}_1+\vec{k}_5-\vec{P}_C)~\psi_{n_C}^{L_Cl_C}(\vec{k}_1,\vec{k}_5)|q_1(\vec{k}_1)\bar{q}_5(\vec{k}_5)\rangle.
\end{split}\label{A3}
\end{equation}
Here, the symbols of $\prod_i~(i=A, B$, and $C)$ represent the Clebsch-Gordan coefficients for the initial and final hadrons, which arise from the couplings among the orbital, spin, and total angular momentum and their projection of $l_z$ and $s_z$ to $j_z$. More specifically, $\prod_i~(i=A, B$, and $C)$ are given by
\begin{equation}
\begin{split}
&\langle s_1m_1, s_2m_2|s_{12}m_{12}\rangle\langle s_{12}m_{12}, s_3m_3|S_As_A\rangle\langle L_Al_A, S_As_A|J_Aj_A\rangle,\\&
\langle s_2m_2, s_5m_5|s_{25}m_{25}\rangle\langle s_{25}m_{25}, s_3m_3|S_Bs_B\rangle\langle L_Bl_B, S_Bs_B|J_Bj_B\rangle,\\&
\langle s_1m_1, s_4m_4|S_Cs_C\rangle\langle L_Cl_C, S_Cs_C|J_Cj_C\rangle,\nonumber
\end{split}
\end{equation}
respectively.

The helicity amplitude $\mathcal {M}^{j_A,j_B,j_C}(q)$ is defined by
\begin{eqnarray}\label{A4}
\langle BC|\mathcal {\hat{T}}|A\rangle=
\delta^3(\vec{P}_A-\vec{P}_B-\vec{P}_C)\mathcal
{M}^{j_A,j_B,j_C}(q),
\end{eqnarray}
where \emph{q} represents the momentum of an outgoing meson in the rest frame of a meson \emph{A}. For comparison with experiments, one obtains the partial wave amplitudes $\mathcal {M}_{LS}(q)$ via the Jacob-Wick formula~\cite{Jacob:1959at}
\begin{equation}
\begin{aligned}\label{A5}
\mathcal {M}_{LS}(q)=&\frac{\sqrt{2L+1}}{2J_A+1}\sum_{\text{$j_B$,$j_C$}}\langle L0Jj_A|J_Aj_A\rangle\\
&\times\langle J_Bj_B,J_Cj_C|Jj_A\rangle\mathcal
{M}^{j_A,j_B,j_C}(q).
\end{aligned}
\end{equation}

Then the decay width $\Gamma(A\rightarrow BC)$ is derived analytically in terms of the partial wave amplitudes in the $A$ rest frame,
\begin{equation}
\begin{aligned}\label{A6}
\Gamma(A\rightarrow BC)=2\pi\frac{E_BE_C}{M_A}q\sum_{L,S}|\mathcal
{M}_{LS}(q)|^2.
\end{aligned}
\end{equation}

Finally, the full expression of $\mathcal {M}_{LS}(q)$ in the rest frame of the baryon \emph{A} is
\begin{widetext}
\begin{equation}
\begin{aligned}
\mathcal {M}_{LS}(q)=&-3\gamma\sum_{l_i,m_j}\langle L0;Jj|J_Aj_A\rangle\langle J_Bj_B;J_Cj_C|Jj\rangle\langle s_1m_1;s_2m_2|s_{dA}m_{12}\rangle\langle s_{dA}m_{12};s_3m_3|S_As_A\rangle\langle S_As_A;L_Al_A|J_Aj_A\rangle\langle s_2m_2;s_5m_5|s_{dB}m_{25}\rangle \\
&\langle s_{dB}m_{25};s_3m_3|S_Bs_B\rangle\langle S_Bs_B;L_Bl_B|J_Bj_B\rangle\langle s_1m_1;s_4m_4|S_Cs_C\rangle\langle S_Cs_C;L_Cl_C|J_Cj_C\rangle\langle s_4m_4;s_5m_5|1-m\rangle\langle1,m;1,-m|0,0\rangle \\
&\langle \varphi^{235}_B\varphi^{14}_C|\varphi^{45}_0\varphi^{123}_A\rangle\langle\omega^{235}_B\omega^{14}_C|\omega^{45}_0\omega^{123}_A\rangle\int\cdots\int d^3\vec{k}_1\cdots d^3\vec{k}_5\delta^3(\vec{k}_1+\vec{k}_2+\vec{k}_3)\delta^3(\vec{q}-\vec{k}_1-\vec{k}_4)\delta^3(\vec{q}+\vec{k}_2+\vec{k}_3+\vec{k}_5)\\
&\delta^3(\vec{k}_4+\vec{k}_5)\Psi_A(\vec{k}_1,\vec{k}_2,\vec{k}_3)\Psi^*_B(\vec{k}_1,\vec{k}_2,\vec{k}_4)\psi^*_C(\vec{k}_3,\vec{k}_5)\mathcal
{Y}_1^m(\frac{\vec{k}_3-\vec{k}_4}{2}),
\label{A7}
\end{aligned}
\end{equation}
\end{widetext}
where, $i=A, B, C$ and $j=1, 2, \cdots, 5$. The color matrix element $\langle\omega^{235}_B\omega^{14}_C|\omega^{45}_0\omega^{123}_A\rangle$ is a constant which can be absorbed into the parameter $\gamma$. The flavor matrix element $\xi=\langle \varphi^{235}_B\varphi^{14}_C|\varphi^{45}_0\varphi^{123}_A\rangle$ will be presented in the next Subsection. To obtain the analytical amplitudes, the SHO wave functions are employed to describe the spatial wave function of a hadron. In the momentum space, the SHO radial wave function, $\psi^n_{Lm}(\textbf{q})$, is given by
\begin{equation}
\begin{split}
&\psi^n_{Lm}(\textbf{q}) \\&=\frac{(-1)^n}{\beta^{3/2}}\sqrt{\frac{2(2n-1)!}{\Gamma(n+L+\frac{1}{2})}}\left(\frac{q}{\beta}\right)^L
e^{-\frac{q^2}{2\beta^2}}L^{L+1/2}_{n-1}\left(\frac{q^2}{\beta^2}\right)\mathcal
{Y}_{Lm}(\textbf{q}),
\end{split}\label{A8}
\end{equation}
with $\textbf{q}=(m_i\vec{k}_j-m_j\vec{k}_i)/(m_i+m_j)$ and $\mathcal {Y}_{Lm}(\textbf{q})=|\textbf{q}|^LY_{Lm}(\Omega_\textbf{q})$. $L^{L+1/2}_{n-1}(q^2/\beta^2)$ is an associated Laguerre polynomial. The values of the SHO wave function scale parameter $\beta$ have been given in Tables \ref{table5} and \ref{table6}. In the light quark cluster picture, the wave function of a charmed baryon can be easily constructed. Taking the \emph{A} baryon as an example, the wave functions corresponding to the $1S$, $2S$, and $1P$ states are given as follows, respectively,
\begin{equation}
\Psi^0_{00}=\frac{3^{3/4}}{\pi^{3/2}\beta_{dA}^{3/2}\beta_{A}^{3/2}}e^{-\frac{1}{2\beta_{dA}^2}\left(\frac{m_1\vec{k}_2-m_2\vec{k}_1}{m_1+m_2}\right)^2-\frac{1}{2\beta_A^2}\left[\frac{(m_1+m_2)\vec{k}_3-m_Q(\vec{k}_1+\vec{k}_2)}{m_1+m_2+m_Q}\right]^2};\nonumber
\end{equation}

\begin{equation}
\begin{split}
\Psi^1_{00}=&-\frac{3^{3/4}}{\sqrt{6}\pi^{3/2}\beta_{dA}^{3/2}\beta_{A}^{3/2}}e^{-\frac{1}{2\beta_{dA}^2}\left(\frac{m_1\vec{k}_2-m_2\vec{k}_1}{m_1+m_2}\right)^2-\frac{1}{2\beta_A^2}\left[\frac{(m_1+m_2)\vec{k}_3-m_Q(\vec{k}_1+\vec{k}_2)}{m_1+m_2+m_Q}\right]^2}\\
&\times\left\{3-\frac{2}{\beta^2_A}\left[\frac{(m_1+m_2)\vec{k}_3-m_Q(\vec{k}_1+\vec{k}_2)}{m_1+m_2+m_Q}\right]^2\right\};\nonumber
\end{split}
\end{equation}

\begin{equation}
\begin{split}
\Psi^0_{1m}=&\frac{3^{3/4}\times2\sqrt{2/3}}{\pi\beta_{dA}^{3/2}\beta_{A}^{5/2}}e^{-\frac{1}{2\beta_{dA}^2}\left(\frac{m_1\vec{k}_2-m_2\vec{k}_1}{m_1+m_2}\right)^2-\frac{1}{2\beta_A^2}\left[\frac{(m_1+m_2)\vec{p}_3-m_Q(\vec{k}_1+\vec{k}_2)}{m_1+m_2+m_Q}\right]^2}\\
&\times\mathcal {Y}_{1m}\left(\frac{(m_1+m_2)\vec{k}_3-m_Q(\vec{k}_1+\vec{k}_2)}{m_1+m_2+m_Q}\right).\nonumber
\end{split}
\end{equation}
With the help of Eq. (\ref{A7}), the transition amplitude can be obtained. In the following, we take the process $\Sigma_c(2520)\rightarrow\Lambda_c(2280)^+\pi$ as an example. The wave functions of initial and final states are
\begin{equation}
\begin{split}
&\Psi_A=\frac{3^{3/4}}{\pi^{3/2}\beta_{dA}^{3/2}\beta_{A}^{3/2}}e^{-\frac{1}{2\beta_{dA}^2}\left(\frac{m_1\vec{k}_2-m_2\vec{k}_1}{m_1+m_2}\right)^2-\frac{1}{2\beta_A^2}\left[\frac{(m_1+m_2)\vec{k}_3-m_3(\vec{k}_1+\vec{k}_2)}{m_1+m_2+m_3}\right]^2};\\&
\Psi_B=\frac{3^{3/4}}{\pi^{3/2}\beta_{dB}^{3/2}\beta_{B}^{3/2}}e^{-\frac{1}{2\beta_{dB}^2}\left(\frac{m_5\vec{k}_2-m_2\vec{k}_5}{m_2+m_5}\right)^2-\frac{1}{2\beta_B^2}\left[\frac{(m_2+m_5)\vec{k}_3-m_Q(\vec{p}_2+\vec{k}_5)}{m_2+m_3+m_5}\right]^2};\\&
\psi_C=-\frac{1}{\pi^{3/4}\beta_C^{3/2}}e^{-\frac{1}{2\beta_C^2}\left(\frac{m_1\vec{k}_4-m_4\vec{k}_1}{m_1+m_2}\right)^2}.\nonumber
\end{split}
\end{equation}
Based on Eq. (\ref{A7}), we obtain the amplitude as
\begin{equation}\label{A9}
\mathcal {M}_{1\frac{1}{2}}(q)=-\frac{3g}{8\pi^{5/4}f^{5/2}\lambda^{3/2}\beta_A^{3/2}\beta_{dA}^{3/2}\beta_B^{3/2}\beta_{dB}^{3/2}\beta_C^{3/2}}pe^{-\frac{4fg-g^2}{4f}q^2}.
\end{equation}
where
\begin{equation}
\begin{split}
&f=\frac{1}{2\beta_{dA}^2}+\frac{1}{2\beta_{dB}^2}+\frac{1}{2\beta_M^2}-\frac{\mu^2}{4\lambda};\\&
g=\frac{1}{\beta_{dA}^2}+\frac{\varepsilon_3}{\beta_{dB}^2}+\frac{\varepsilon_4}{\beta_M^2}-\frac{\mu\nu}{2\lambda};\\&
h=\frac{1}{2\beta_{dA}^2}+\frac{\varepsilon_2^2}{2\beta_B^2}+\frac{\varepsilon_3^2}{2\beta_{dB}^2}+\frac{\varepsilon_4^2}{2\beta_M^2}-\frac{\nu^2}{4\lambda};\\&
\lambda=\frac{1}{2\beta_A^2}+\frac{1}{2\beta_B^2}+\frac{\varepsilon_1^2}{2\beta_{dA}^2}+\frac{\varepsilon_3^2}{2\beta_{dB}^2};\\&
\mu=\frac{\varepsilon_1}{\beta_{dA}^2}+\frac{\varepsilon_3}{\beta_{dB}^2};~~\nu=\frac{\varepsilon_1}{\beta_{dA}^2}+\frac{\varepsilon_3}{\beta_B^2}+\frac{\varepsilon_3^2}{\beta_{dB}^2};\\&
\varepsilon_1=\frac{m_1}{m_1+m_2};~~\varepsilon_2=\frac{m_3}{m_1+m_3+m_5};\\&
\varepsilon_3=\frac{m_5}{m_2+m_5};~~\varepsilon_4=\frac{m_4}{m_1+m_4},\nonumber
\end{split}
\end{equation}
and
\begin{equation}\label{A10}
q=\frac{\sqrt{[M_A^2-(M_B+M_C)^2][M_A^2-(M_B-M_C)^2]}}{2M_A}.
\end{equation}
Here, $M_A$, $M_B$, and $M_C$ are the masses of hadrons \emph{A}, \emph{B}, and \emph{C}, respectively. Then the $\tilde{\beta}$ in Eq. (\ref{eq20}) is given by
\begin{equation}\label{A11}
\tilde{\beta}=2\sqrt{\frac{f}{4fg-g^2}},
\end{equation}
where \emph{f} and \emph{g} have been defined above. For the decay channel of $\Sigma_c(2520)\rightarrow\Lambda_c(2280)\pi$, the value of $\mathcal {C}^{s_Q,j_B,J_B}_{j_C,j_A,J_A}$ is $-1$. Therefore, we obtain
\begin{equation}\label{A12}
\mathcal {M}^{1,0}_{11}(q)=\frac{3g}{8\pi^{5/4}f^{5/2}\lambda^{3/2}\beta_A^{3/2}\beta_{dA}^{3/2}\beta_B^{3/2}\beta_{dB}^{3/2}\beta_C^{3/2}},
\end{equation}
where a phase space factor $(2\pi E_BE_C/M_A)^{1/2}\gamma$ is omitted. One notices that the unitary rotation between the $LS$ coupling and $jj$ coupling (Eq. (\ref{eq15})) should be performed to reduce the transition factors of 1\emph{P} state with the same $J^P$. More details for calculating the decay amplitudes of an excited baryon in the $^3P_0$ model can be found in the Refs.~\cite{Chen:2007xf,Capstick:1992th}.

\section{Flavor factors}\label{2}
\begin{table}
\caption{The flavor matrix element $\xi$ for different decay channels of the charmed baryons.} \label{table13}
\renewcommand\arraystretch{1.2}
\begin{tabular*}{85mm}{@{\extracolsep{\fill}}cccccc }
\toprule[1pt]\toprule[1pt]
Initial state  &\multicolumn{5}{c}{Final states}   \\
\cline{1-1}\cline{2-6}
 $\Lambda_c^+$      & $\Sigma_c^{++,+,0}\pi^{-,0,+}$     & $D^+n/D^0p$               &                         &                   &       \\
                    &  $\sqrt{1/3}$                      & $\sqrt{1/3}$              &                         &                   &      \\
$\Sigma_c^{++}$     & $\Sigma_c^{++,+}\pi^{0,+}$         & $\Lambda_c^+\pi^+$        &    $D^+p$               &                   &      \\
                    &  $\sqrt{1/3}$                      & $\sqrt{1/3}$              &  $\sqrt{1/6}$           &                   &      \\
$\Sigma_c^{+}$      & $\Sigma_c^{++,0}\pi^{-,+}$         & $\Lambda_c^+\pi^0$        &    $D^+n/D^0p$          &                   &       \\
                    &  $\sqrt{1/3}$                      & $-\sqrt{1/3}$             &  $\sqrt{1/12}$          &                   &      \\
$\Sigma_c^0$        & $\Sigma_c^+\pi^-/\Sigma_c^0\pi^0$  & $\Lambda_c^+\pi^-$        &    $D^0n$               &                   &      \\
                    &  $\sqrt{1/3}$                      & $\sqrt{1/3}$              &  $\sqrt{1/6}$           &                   &      \\
$\Xi_c^{(\prime)+}$ & $\Lambda_c^+K^0$                   & $\Xi_c^{(\prime)0}\pi^+$  &$\Xi_c^{(\prime)+}\pi^0$ &$\Sigma_c^{++}K^-$ & $\Sigma_c^+K^0$ \\
                    &  $\sqrt{1/6}$                      & $\sqrt{1/6}$              &  $\sqrt{1/12}$          &   $\sqrt{1/3}$    & $\sqrt{1/6}$ \\
$\Xi_c^{(\prime)0}$ & $\Lambda_c^+K^-$                   & $\Xi_c^{(\prime)+}\pi^-$  &$\Xi_c^{(\prime)0}\pi^0$ &$\Sigma_c^+K^-$    & $\Sigma_c^0K^0$ \\
                    &  $\sqrt{1/6}$                      & $\sqrt{1/6}$              &  $\sqrt{1/12}$          & $\sqrt{1/6}$      & $\sqrt{1/3}$     \\
\bottomrule[1pt]\bottomrule[1pt]
\end{tabular*}
\end{table}

Based on the light SU(3) flavor symmetry, the flavor wave functions of charmed and charmed-strange baryons are given by~\cite{Roberts:2007ni}
\begin{equation}
\begin{split}
&\Lambda_c^+=\frac{1}{\sqrt{2}}(ud-du)c;~~~~\Sigma_c^{++}=uuc;\\&
\Xi_c^+=\frac{1}{\sqrt{2}}(us-su)c;~~~~~\Sigma_c^+=\frac{1}{\sqrt{2}}(ud+du)c;\\&
\Xi_c^0=\frac{1}{\sqrt{2}}(ds-sd)c;~~~~~\Sigma_c^0=ddc;\\&
\Xi^{'+}_c=\frac{1}{\sqrt{2}}(us+su)c;~~~~\Xi^{'0}_c=\frac{1}{\sqrt{2}}(ds+sd)c.\nonumber
\end{split}
\end{equation}
As shown in Fig.\ref{Fig3}, the final states of an excited charmed baryons may contain a light meson and a low energy charmed baryon or a light baryon and a charmed meson. The flavor wave functions for the final states are collected in the following
\begin{equation}
\begin{split}
&\pi^+=u\bar{d};~~~~\pi^-=d\bar{u};~~~~\pi^0=(u\bar{u}-d\bar{d})/\sqrt{2};\\&
K^-=\bar{u}s;~~~~~\bar{K}^0=\bar{d}s;~~~~~D^+=\bar{d}c;~~~~~D^0=\bar{u}c;\\&
p=\frac{1}{\sqrt{2}}(du-ud)u;~~~~~n=\frac{1}{\sqrt{2}}(du-ud)d;\\&
\Lambda^0=\frac{1}{\sqrt{2}}(du-ud)s.\nonumber
\end{split}
\end{equation}
With the above flavor wave functions, the flavor matrix elements $\xi$ for different decay processes are presented in Table~\ref{table13}.


\end{document}